\pdfoutput=1
\documentclass[3p,12pt]{elsarticle}

\usepackage[british]{babel}
\usepackage{amsmath,amssymb,amsthm}
\usepackage{graphicx}
\usepackage{booktabs}
\usepackage{array}
\usepackage{longtable}
\usepackage{microtype}
\newif\iflinenumbers \linenumbersfalse
\usepackage{lineno}
\usepackage{pdfpages}
\usepackage[hidelinks,hypertexnames=false]{hyperref}
\usepackage[capitalise,nameinlink]{cleveref}
\biboptions{sort&compress}
\hypersetup{pdftitle={The norm, not the labels: training and evaluating finite-element
  surrogates for stress analysis in the energy norm}}
\hypersetup{pdfauthor={Ruifeng Cao, Xidan Song}}
\makeatletter
\g@addto@macro{\UrlBreaks}{\do\_\do\.}
\myfooterfont{\scriptsize\itshape}
\def\ps@pprintTitle{\let\@oddhead\@empty\let\@evenhead\@empty
  \def\@oddfoot{\@myfooterfont Preprint\hfill\@date}%
  \let\@evenfoot\@oddfoot}
\makeatother

\date{\number\day\space\ifcase\month\or January\or February\or March\or April\or May\or
  June\or July\or August\or September\or October\or November\or December\fi
  \space\number\year}

\newtheorem{lemma}{Lemma}
\newtheorem{proposition}{Proposition}
\newtheorem{corollary}{Corollary}
\theoremstyle{remark}
\newtheorem{remark}{Remark}
\crefname{lemma}{Lemma}{Lemmas}
\crefname{proposition}{Proposition}{Propositions}
\crefname{corollary}{Corollary}{Corollaries}
\crefname{remark}{Remark}{Remarks}

\newcommand{\norm}[1]{\lVert #1\rVert}
\newcommand{\gap}{\operatorname{gap}}
\newcommand{\vm}{\sigma_{\mathrm{vm}}}
\newcommand{\Ustar}{U^{\ast}}
\newcommand{\cstar}{c^{\ast}}
\newcommand{\R}{\mathbb{R}}

\newcommand{\figfull}[1]{\makebox[\linewidth][c]{%
  \includegraphics[width=\dimexpr\paperwidth-4cm\relax]{#1}}}

\newcommand{\numCmAdvMax}{99.8\%}
\newcommand{\numCmAdvMin}{71.3\%}

\newcommand{\numCmFreeDispExcessAbs}{25.6\%}

\newcommand{\numCmFreeGapSeeds}{0.0374, 0.0405 and 0.0371}

\newcommand{\numCmFreeVm}{0.123}

\newcommand{\numCmKnormDisp}{0.0547}
\newcommand{\numCmKnormGap}{0.0290}

\newcommand{\numCmKnormGapSeeds}{0.0370, 0.0146 and 0.0355}

\newcommand{\numCmKnormVm}{0.106}

\newcommand{\numCmKnormWorseSixteen}{11}
\newcommand{\numCmKnormWorseSixtyFour}{1}
\newcommand{\numCmKnormWorseTwoFiveSix}{0}
\newcommand{\numCmLabDisp}{0.0505}

\newcommand{\numCmLabOverFreeGap}{3.5}
\newcommand{\numCmLabOverFreeGapRoot}{1.9}

\newcommand{\numCmLabVm}{0.235}

\newcommand{\numCmLabWorseSixteen}{768}
\newcommand{\numCmLabWorseSixtyFour}{532}
\newcommand{\numCmLabWorseTwoFiveSix}{53}

\newcommand{\numCmMgnDisp}{0.0499}

\newcommand{\numCmMgnOverFreeVm}{4.7}
\newcommand{\numCmMgnOverLabGap}{6.2}

\newcommand{\numCmMgnWorse}{187}

\newcommand{\numCmPairs}{768}

\newcommand{\numCmSolves}{5,120}
\newcommand{\numCmSolvesTrain}{4,096}

\newcommand{\numCoarsenHigh}{2.5}

\newcommand{\numCoarsenLow}{1.8}

\newcommand{\numCostAsmOverRitzMax}{27}
\newcommand{\numCostAsmOverRitzMin}{6}
\newcommand{\numCostAsmVal}{900 ms}
\newcommand{\numCostBreakEvenDirect}{84,000}
\newcommand{\numCostBreakEvenMatch}{400,000}

\newcommand{\numCostFineDirectSolved}{seven}

\newcommand{\numCostLabelMin}{21}

\newcommand{\numCostRitzValMax}{144 ms}
\newcommand{\numCostRitzValMin}{34 ms}

\newcommand{\numCostSurrVal}{49 ms}
\newcommand{\numCostThreads}{25}
\newcommand{\numCostThreeFineDirectRatio}{64}
\newcommand{\numCostThreeFineGap}{0.12}
\newcommand{\numCostThreeFineMatchRatio}{1.9}
\newcommand{\numCostThreeValDirectRatio}{27}
\newcommand{\numCostThreeValGap}{0.0085}
\newcommand{\numCostThreeValMatchRatio}{5.2}
\newcommand{\numCostTwoDirectRatio}{1.2}

\newcommand{\numEtwoNodesFine}{41,367}

\newcommand{\numEtwoTransformerFineStep}{14.8}

\newcommand{\numFieldLoadWorseLab}{126}
\newcommand{\numFieldLoadWorseMgn}{106}
\newcommand{\numFieldLoads}{3,072}
\newcommand{\numFieldRqAboveLab}{89.0\%}

\newcommand{\numFieldRqFree}{11.9}
\newcommand{\numFieldRqLab}{132}
\newcommand{\numFieldRqMgn}{1,026}

\newcommand{\numFigWorstFreeC}{0.78}
\newcommand{\numFigWorstFreeDisp}{0.22}
\newcommand{\numFigWorstFreeDispC}{0.045}
\newcommand{\numFigWorstLoad}{downward body force}
\newcommand{\numFigWorstTwiceFree}{12\%}
\newcommand{\numFigWorstTwiceLab}{67\%}
\newcommand{\numFigWorstTwiceMgn}{52\%}

\newcommand{\numGTwoAdvMin}{98.9\%}

\newcommand{\numGTwoParity}{$-$36.3\%}

\newcommand{\numHOneRel}{71.3\%}
\newcommand{\numHOneTau}{12.4\%}

\newcommand{\numHThreeRel}{24.2\%}
\newcommand{\numHThreeResHigh}{28.0\%}
\newcommand{\numHThreeResLow}{21.9\%}
\newcommand{\numHThreeSeedRel}{1\%, 64\% and 4\%}

\newcommand{\numHThreeTau}{38.1\%}
\newcommand{\numHTwoARel}{78.2\%}
\newcommand{\numHTwoARootRel}{53.4\%}
\newcommand{\numHTwoATau}{16.4\%}
\newcommand{\numHTwoBRel}{54.9\%}
\newcommand{\numHTwoBTau}{12.3\%}

\newcommand{\numHoldHTwoABaseSeeds}{0.142, 0.147 and 0.121}
\newcommand{\numHoldHTwoAMedHold}{0.0223}
\newcommand{\numHoldHTwoAMedRel}{81.4\%}
\newcommand{\numHoldHTwoAMedTau}{13.4\%}
\newcommand{\numHoldHTwoAMedVal}{0.0210}
\newcommand{\numHoldHTwoANewSeeds}{0.130, 0.0213 and 0.0834}
\newcommand{\numHoldHTwoARatio}{2.69}
\newcommand{\numHoldHTwoARatioHigh}{4.64}
\newcommand{\numHoldHTwoARatioLow}{1.12}
\newcommand{\numHoldHTwoARel}{42.9\%}

\newcommand{\numHoldHTwoATau}{47.4\%}

\newcommand{\numHoldHTwoATopSeeds}{254.3, 21.6 and 190.4}

\newcommand{\numHoldTThreeLab}{340}
\newcommand{\numHoldTThreeLoads}{8,232}
\newcommand{\numHoldTThreeMgn}{237}

\newcommand{\numHoldThreeN}{686}

\newcommand{\numHoldThreeTail}{720}

\newcommand{\numHoldTwoFreeGap}{0.0393}
\newcommand{\numHoldTwoKnormGap}{0.0781}

\newcommand{\numHoldTwoN}{4,111}
\newcommand{\numHoldTwoPairs}{12,333}

\newcommand{\numHoldTwoTail}{4,144}
\newcommand{\numHoldTwoWorseInstFree}{none}
\newcommand{\numHoldTwoWorseInstKnorm}{9}
\newcommand{\numHoldTwoWorseInstLab}{16}

\newcommand{\numHoldVAlpha}{5\%}
\newcommand{\numHoldVQuantities}{35}

\newcommand{\numKappaThreeHigh}{\ensuremath{5.2 \times 10^{5}}}
\newcommand{\numKappaThreeLow}{\ensuremath{2.9 \times 10^{4}}}
\newcommand{\numKappaTwoHigh}{\ensuremath{1.5 \times 10^{5}}}
\newcommand{\numKappaTwoLow}{\ensuremath{4.8 \times 10^{3}}}

\newcommand{\numKdFreeDispRel}{25.0\%}

\newcommand{\numKdFreeMedMax}{0.0093}
\newcommand{\numKdFreeMedMin}{0.0058}
\newcommand{\numKdFreePairs}{738}
\newcommand{\numKdFreePeakRel}{51.7\%}
\newcommand{\numKdFreeRel}{41.0\%}
\newcommand{\numKdFreeTau}{27.0\%}
\newcommand{\numKdFreeVmRel}{19.4\%}
\newcommand{\numKdFreeWelchHigh}{+16.7\%}
\newcommand{\numKdFreeWelchLow}{$-$98.7\%}
\newcommand{\numKdGapRel}{98.2\%}
\newcommand{\numKdGapTau}{11.7\%}
\newcommand{\numKdHoldFreeRel}{38.0\%}
\newcommand{\numKdHoldFreeTau}{25.6\%}
\newcommand{\numKdHoldGapRel}{98.0\%}
\newcommand{\numKdHoldGapTau}{12.6\%}
\newcommand{\numKdHoldVmRel}{83.0\%}
\newcommand{\numKdHoldVmTau}{10.0\%}
\newcommand{\numKdKnormGapMedian}{0.0044}
\newcommand{\numKdLabGapMedian}{0.25}
\newcommand{\numKdMeansGapRel}{99.98\%}
\newcommand{\numKdMeansGapTau}{197.9\%}
\newcommand{\numKdMeansVmRel}{88.1\%}
\newcommand{\numKdMeansVmTau}{43.7\%}
\newcommand{\numKdMeansVmWelchHigh}{+5.8\%}
\newcommand{\numKdMeansVmWelchLow}{$-$182.0\%}
\newcommand{\numKdSecondary}{30}
\newcommand{\numKdVmRel}{83.6\%}
\newcommand{\numKdVmTau}{10.0\%}

\newcommand{\numMgnDispLowerMax}{92.6\%}
\newcommand{\numMgnDispLowerMin}{84.4\%}
\newcommand{\numMgnGapHigherMin}{99.2\%}
\newcommand{\numMgnLowerDispHigherVm}{685}
\newcommand{\numMgnPeakLowerMax}{29\%}
\newcommand{\numMgnPeakLowerMin}{17\%}
\newcommand{\numMgnVmHigher}{767}

\newcommand{\numNlCmLabelE}{1.14}
\newcommand{\numNlCmLabelVm}{1.12}
\newcommand{\numNlCmNormE}{2.5}
\newcommand{\numNlCmNormEMean}{2.5}
\newcommand{\numNlCmNormVm}{2.2}
\newcommand{\numNlCmNormVmMean}{2.2}

\newcommand{\numNlThreeLabelE}{1.30}
\newcommand{\numNlThreeLabelVm}{1.24}
\newcommand{\numNlThreeLabelWithin}{1.3}
\newcommand{\numNlThreeNormE}{7.5}
\newcommand{\numNlThreeNormEMean}{10.7}
\newcommand{\numNlThreeNormVm}{6.1}
\newcommand{\numNlThreeNormVmMean}{8.4}
\newcommand{\numNlThreeNormVmWord}{six}

\newcommand{\numNodesFineMax}{71,749}
\newcommand{\numNodesFineMedian}{36,216}
\newcommand{\numNodesFineMin}{15,175}
\newcommand{\numNodesInMax}{17,692}
\newcommand{\numNodesInMedian}{5,617}
\newcommand{\numNodesInMin}{2,195}

\newcommand{\numPairCmKnormLabDisp}{49.7\%}
\newcommand{\numPairCmKnormLabGap}{99.9\%}

\newcommand{\numPairCmKnormLabVm}{99.9\%}
\newcommand{\numPairCmLabDisp}{29.9\%}
\newcommand{\numPairCmLabGap}{99.9\%}

\newcommand{\numPairCmLabVm}{99.9\%}

\newcommand{\numPairCmOpposite}{69.9\%}

\newcommand{\numPairThreeLabDisp}{78.9\%}
\newcommand{\numPairThreeLabGap}{99.9\%}

\newcommand{\numPairThreeLabVm}{99.9\%}

\newcommand{\numParamsGraph}{3.9}
\newcommand{\numParamsTransformer}{6.7}

\newcommand{\numRitzHoldKnormGapMin}{88\%}
\newcommand{\numRitzHoldKnormPeakMiss}{two}
\newcommand{\numRitzThreeK}{168, 252, 360 and 495}
\newcommand{\numRitzThreeKMax}{495}
\newcommand{\numRitzTwoFreeDispMax}{97\%}
\newcommand{\numRitzTwoFreeDispMin}{80\%}
\newcommand{\numRitzTwoFreeGapEightMax}{66\%}
\newcommand{\numRitzTwoFreeGapEightMin}{61\%}
\newcommand{\numRitzTwoFreeGapSixMax}{20\%}
\newcommand{\numRitzTwoFreeGapSixMin}{16\%}
\newcommand{\numRitzTwoFreePeakMin}{96.8\%}
\newcommand{\numRitzTwoK}{42 and 72}
\newcommand{\numRitzTwoKEight}{72}
\newcommand{\numRitzTwoKSix}{42}
\newcommand{\numRitzValFreeDispNineMax}{100\%}
\newcommand{\numRitzValFreeDispNineMin}{99.6\%}
\newcommand{\numRitzValFreeDispSixMax}{79\%}
\newcommand{\numRitzValFreeDispSixMin}{67\%}
\newcommand{\numRitzValFreePeakMin}{98.4\%}
\newcommand{\numRitzValGapNine}{0.0063}
\newcommand{\numRitzValGapSix}{0.018}
\newcommand{\numRitzValKnormGapMin}{89.8\%}
\newcommand{\numRitzValNineDispAll}{97.6\%}
\newcommand{\numRitzValPeakNine}{0.23}
\newcommand{\numRitzValPeakRatioMax}{4.9}
\newcommand{\numRitzValPeakRatioMin}{2.4}
\newcommand{\numRitzValPeakSix}{0.32}

\newcommand{\numSpecLabKnormAbove}{94.9\%}

\newcommand{\numSpecLabKnormDisp}{1.26}

\newcommand{\numSpecLabKnormGap}{9.85}
\newcommand{\numSpecLabKnormRq}{7.80}
\newcommand{\numSpecLabKnormRqSeeds}{8.26, 9.46 and 6.07}

\newcommand{\numSpecLoadWorseMgn}{724}
\newcommand{\numSpecLoads}{3,072}

\newcommand{\numSpecMgnLabDisp}{1.00}

\newcommand{\numSpecMgnLabRq}{4.33}
\newcommand{\numSpecRqFree}{10.3}
\newcommand{\numSpecRqKnorm}{11.3}
\newcommand{\numSpecRqLab}{115}
\newcommand{\numSpecRqMgn}{553}
\newcommand{\numSpecRqStarOverMin}{1.27}

\newcommand{\numSpecShareKFree}{69\%}
\newcommand{\numSpecShareKKnorm}{73\%}
\newcommand{\numSpecShareKLab}{95\%}

\newcommand{\numSpecShareKStar}{1.5\%}
\newcommand{\numSpecSpectralShare}{90\%}

\newcommand{\numSpecVmAreaRatio}{2.26}

\newcommand{\numSpecVmElemRatio}{2.22}
\newcommand{\numSqrtKappaMax}{723}
\newcommand{\numSqrtKappaMin}{69}

\newcommand{\numTailTwoKnormOther}{0.304}

\newcommand{\numThreeAncOverFreeGap}{12.7}
\newcommand{\numThreeAncOverFreeGapRoot}{3.6}

\newcommand{\numThreeAncVm}{0.128}
\newcommand{\numThreeAncWorse}{2}

\newcommand{\numThreeEvalSolves}{2,048}
\newcommand{\numThreeFewSolves}{256}

\newcommand{\numThreeFreeDispAdvantage}{36.3\%}

\newcommand{\numThreeFreeGapMedian}{0.00763}
\newcommand{\numThreeFreeLoadWorse}{1}

\newcommand{\numThreeFreeOverMgnDisp}{1.8}

\newcommand{\numThreeFreeVm}{0.0554}

\newcommand{\numThreeGuardTotal}{ten}
\newcommand{\numThreeGuardWithin}{three}
\newcommand{\numThreeKnormDisp}{0.0227}

\newcommand{\numThreeLabDisp}{0.0470}

\newcommand{\numThreeLabGapMedian}{0.252}

\newcommand{\numThreeLabVm}{0.370}
\newcommand{\numThreeLabWorse}{18}

\newcommand{\numThreeLabWorst}{23,727}
\newcommand{\numThreeLabWorstSeed}{1}
\newcommand{\numThreeMedianRatioMax}{33}
\newcommand{\numThreeMedianRatioMin}{14}
\newcommand{\numThreeMgnDisp}{0.0168}

\newcommand{\numThreeMgnOverFreePeak}{6.0}
\newcommand{\numThreeMgnOverFreeVm}{5.3}

\newcommand{\numThreeMgnWorse}{28}
\newcommand{\numThreePairs}{768}
\newcommand{\numThreeSolves}{6,400}
\newcommand{\numThreeTrainSolves}{4,096}

\newcommand{\numTimeBottleneckSpeedup}{11}
\newcommand{\numTimeThreeSeed}{27.3}

\newcommand{\numWorstFreeDisp}{0.071}
\newcommand{\numWorstFreeGap}{0.0349}
\newcommand{\numWorstFreeVm}{0.10}
\newcommand{\numWorstIndex}{107}
\newcommand{\numWorstLabDisp}{0.032}
\newcommand{\numWorstLabGap}{2.90}
\newcommand{\numWorstLabVm}{0.73}
\newcommand{\numWorstMgnDisp}{0.033}
\newcommand{\numWorstMgnGap}{27.0}
\newcommand{\numWorstMgnVm}{2.4}

\newcommand{\SuppPrereg}{S1}
\newcommand{\SuppThree}{S2}
\newcommand{\SuppTwo}{S3}
\newcommand{\SuppTransfer}{S4}
\newcommand{\SuppWarm}{S5}
\newcommand{\SuppSide}{S6}
\newcommand{\SuppJuly}{S7}
\newcommand{\SuppRepro}{S8}
\newcommand{\SuppRitz}{S9}
\newcommand{\SuppTiming}{S10}

\begin{document}

\begin{frontmatter}

\title{The norm, not the labels: training and evaluating finite-element surrogates for
stress analysis in the energy norm}

\author[man]{Ruifeng Cao\corref{cor1}}
\ead{ruifeng.cao@manchester.ac.uk}
\author[whu]{Xidan Song}
\ead{liuqingdamafen@whu.edu.cn}
\cortext[cor1]{Corresponding author.}

\affiliation[man]{organization={School of Engineering, The University of Manchester},
  addressline={Oxford Road}, city={Manchester}, postcode={M13~9PL},
  country={United Kingdom}}
\affiliation[whu]{organization={School of Computer Science, Wuhan University},
  addressline={299 Bayi Road, Wuchang District}, city={Wuhan}, postcode={430072},
  country={China}}

\begin{abstract}
Finite-element surrogates predict the stresses of a new design in one evaluation, but are
usually trained on solutions with a displacement loss and compared by it, although
engineering decisions rest on stresses. Application in engineering: stress analysis of
perforated plates and solids with cavities, parametric families in which holes concentrate
stress. Contribution to artificial intelligence: training and evaluating such surrogates in
the energy norm, with or without solutions, and screening their predictions without
solutions. For linear elastostatics, minimising the assembled discrete potential energy is
exactly supervised regression in this norm: the energy gap of a prediction is the strain
energy of its stress error. A displacement loss under-weights the stiff modes that carry the
strains. In pre-registered comparisons, changing only the norm of the
supervised loss cut a three-dimensional transformer's median energy-norm error
\numNlThreeNormE{}-fold and its median von Mises error \numNlThreeNormVm{}-fold; in two
dimensions it more than halved the von Mises error, also on held-out instances. In three
dimensions, the transformer trained on the energy alone came within a factor of
\numNlThreeLabelWithin{} of the energy-norm-supervised one in median energy-norm and von
Mises error. It was more accurate than the displacement-supervised one on all five metrics
in every seed. Given the stiffness matrix, a Ritz projection was more accurate than the
label-free transformer, in the seed means, on every metric but peak stress. No network
transferred to a substantially finer mesh. The energy ranks predictions of a load case, and
its sign flags those worse than the zero field.
\end{abstract}


\end{frontmatter}

\iflinenumbers\linenumbers\fi

\section{Introduction}\label{sec:intro}

Holes, cut-outs and internal cavities concentrate stress, and components that contain them are
analysed on their stresses. Such analyses are often repeated many times: in design studies
that vary the number, size and position of the holes, in uncertainty quantification over their
scatter, and in the assessment of parts whose voids differ from one part to the next. Each analysis
is a finite-element solve. The examples here are dimensionless families of perforated plates
in plane stress and of solids with spherical cavities, clamped on one face and carrying four
load cases, with up to three holes or cavities of random size and position
(\cref{sec:methods:data}).

Neural surrogates of finite-element models return, in one network evaluation, the field that
an assembly and a solve would return, so that such studies can query many configurations
cheaply
\cite{Pfaff2021,Lu2021DeepONet,Li2021FNO,He2024GeomDeepONet,Zeng2025PCNO,Liu2026GINOT,Deshpande2024MAgNET};
some predict the stresses directly \cite{JiangChen2023GNN}. They
are usually trained by minimising a mean-square or relative error of the displacement
against reference solutions, and compared by the same error. Both choices fix a norm, the
Euclidean norm of the nodal displacements, and neither is usually discussed. The stresses,
on which engineering decisions depend, are measured by another norm. This paper shows that,
for the surrogates we trained, the choice of norm decided their stress accuracy more than
whether solutions were used at all.

The clearest symptom is a reversal. A transformer trained on the discrete potential energy
of 1,024 three-dimensional instances, without a single solve, had lower displacement, energy
and stress errors in each of three seeds than the same transformer trained on the exact
solutions of the same instances. Some of the supervised transformer's predictions were worse
than the zero field.
Reported comparisons of energy-trained and supervised networks disagree: the two were about
as accurate in one study \cite{Zhu2019}, the energy-trained network was the more accurate in
energy, strain or stress in two others \cite{Dalton2023,MonteiroFernandes2025}, and the less
accurate in a fourth \cite{Sun2026FEVessel}.

The explanation is a classical identity. The energy of a prediction exceeds that of the
solution by half its squared error in the energy norm \cite{Ciarlet1978}, which is the
strain energy of its stress error. Minimising the energy over a training set is therefore
exactly supervised regression in the energy norm, with the same gradients, although no
solution is computed. Minimising the displacement error is regression in the Euclidean norm,
which weights all eigenmodes of the stiffness matrix equally and so under-weights the stiff
modes that carry the strains; a Euclidean residual loss regresses in the norm of the squared
stiffness matrix. Supervision and energy training thus differ, before anything else, in the
norm they minimise.

\begin{figure}[tbp]
\centering
\figfull{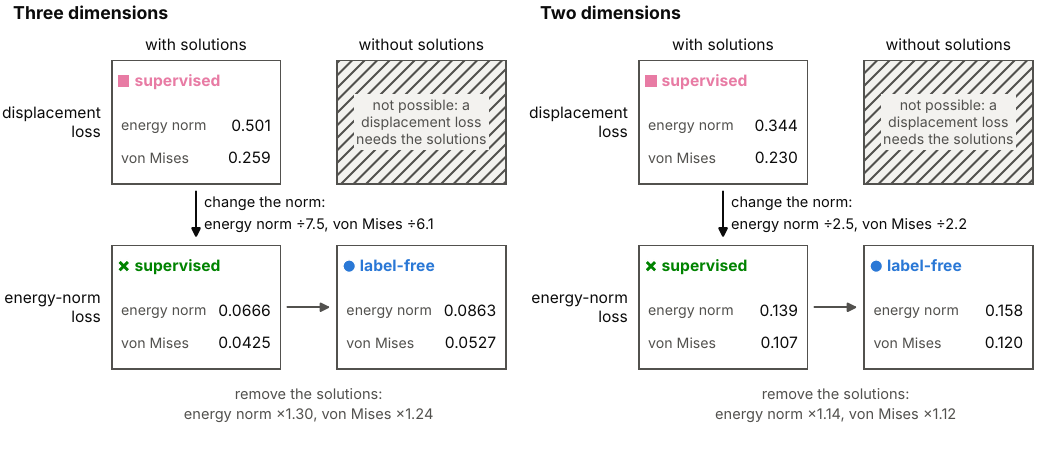}
\caption{The norm of the loss against the labels, in three dimensions (left) and two (right).
The same transformer is trained on the solutions of 1,024 instances with the relative
displacement error (top left) or the relative energy-norm error (bottom left), and on their energy without the solutions (bottom right). Label-free training also changes the weighting of the load cases and the learning rate (\cref{sec:labelfree}); a displacement
loss needs the solutions (top right). Each cell: relative energy-norm and von Mises errors, medians over the 256
validation instances averaged over three seeds.}
\label{fig:normlabels}
\end{figure}

We test this by separating the norm from the labels (\cref{fig:normlabels}). The same
network, on the same instances, is trained on the solutions in the Euclidean norm, on the
solutions in the energy norm, and without solutions in the energy norm; the fourth
combination does not exist, because a displacement loss needs the solutions. Removing the
solutions also changes the weighting of the load cases and the learning rate
(\cref{sec:labelfree}). In three dimensions, changing the norm with the labels kept cut the
median energy-norm error \numNlThreeNormE{}-fold and the median von Mises error
\numNlThreeNormVm{}-fold; removing the labels with the energy norm kept raised them by
factors of \numNlThreeLabelE{} and \numNlThreeLabelVm{}. In two dimensions the energy norm
more than halved the von Mises error of the supervised network. On the means over the instances,
changing the norm divided the energy-norm and von Mises errors by \numNlThreeNormEMean{} and
\numNlThreeNormVmMean{} in three dimensions and by \numNlCmNormEMean{} and
\numNlCmNormVmMean{} in two. The comparisons were pre-registered (\cref{tab:claims}), and
their claims were read again on held-out instances.

The norm matters for evaluation as well. Ranked by displacement error, a graph network
trained on the solutions was the most accurate of the three-dimensional networks; ranked by
von Mises error, it was among the least accurate, at \numThreeMgnOverFreeVm{} times the
label-free transformer's. Predictions can also be checked with the energy, without a solution:
differences of energy rank the predictions of a load case by their energy-norm errors, and
the sign of the energy shows whether a prediction is worse than the zero field. A posteriori
error estimators bound the energy-norm error itself, at the price of an equilibrated stress
field or a few Krylov iterations (\cref{sec:related}); these two checks need neither.

The networks did not outperform classical methods on these problems. Given the stiffness matrix, a
polynomial Ritz projection with \numRitzThreeKMax{} unknowns was more accurate than the
label-free transformer, in the seed means, on every metric but peak stress. Conjugate
gradients, stopped by a test that uses the reference solution, reached a network's accuracy within a
few times its inference time (\cref{sec:cost}). Linear
elastostatics is the setting in which the identity
is exact and the comparison can be kept narrow. Its lesson, the choice of norm, should
matter wherever surrogates are trained and compared; training without solutions is most useful
where solves are expensive.

The contributions are three. First, an exact account of what three common losses regress on the assembled system
(\cref{sec:theory}). The energy regresses in the energy norm, which bounds the von Mises error
for a homogeneous isotropic material; the displacement error, in the Euclidean norm; and the
Euclidean residual, in the norm of the squared stiffness matrix. A
spectral factor relates the energy-norm and the displacement errors and shows when they rank
predictions in opposite orders. Second, a controlled separation of the norm of the loss from
the labels in two and three dimensions (\cref{sec:norm,sec:labelfree}), with criteria fixed
before the runs and read again on held-out instances (\cref{sec:holdout}), where seven of
eight claims replicated and one, the two-dimensional reduction of the mean relative energy
gap, did not. Third, evidence that displacement errors misrank surrogates by stress, and two checks
that need no solution, with a closed-form rescaling for predictions that fail them
(\cref{sec:checks}). A classical baseline and the cost of each step (\cref{sec:cost}) bound
these claims. \Cref{lem:exact}, \cref{rem:norms} and two statements on conditioning and
on gradient descent are restated from a companion note \cite{CaoSong2026}; supplementary
Section~\SuppRepro{} gives the statements on conditioning and on gradient descent with their
checks. That note checks the restated results on synthetic
problems and on probe instances; the trained networks, their comparisons and the baselines are
new here.

For engineering use, the results suggest three rules, which \cref{sec:discussion:engineering}
states with their costs and limits. Train a surrogate used for stress analysis in the energy
norm, on the solutions where they have been computed and on the energy alone where they have
not. Compare surrogates by their energy-norm and stress errors, not by their displacement
errors alone. Screen each prediction by the sign of its energy before it is used, and send
those that fail back to the solver. The screen catches gross failures only.
\section{Related work}\label{sec:related}

\subsection{Energy and variational training of neural solvers}
The deep Ritz method \cite{EYu2018} minimises the continuous energy of a variational problem
over neural trial functions, and the deep energy method \cite{Samaniego2020} brought the idea to computational mechanics. Later work treated essential boundary conditions
\cite{LiaoMing2021} and the error of the quadrature that evaluates the continuous functional
\cite{LiuCaiRamani2023}, and variational and weak-form losses followed
\cite{Kharazmi2021,Wang2022CENN,Wang2025KINN,Manav2024}. Physics-informed neural networks
\cite{Raissi2019} minimise instead a Euclidean residual of the differential equation; for
two- and three-dimensional solid mechanics, a least-squares weighted residual loss has been
compared with this collocation loss and with the energy \cite{Bai2023LSWR}. Robust
variational physics-informed networks \cite{Rojas2024} and the deep Fourier residual method
\cite{Taylor2023DFR,Taylor2024} construct losses equivalent to the energy norm of the error by measuring the residual in a
dual norm; the discrete form of that norm is the energy norm of \cref{sec:theory}.
\Cref{lem:exact} and \cref{rem:norms} are the exact algebraic counterpart of that
equivalence on the assembled system, without quadrature error or equivalence constants,
because the inverse in the dual norm cancels. Energy natural gradient descent trains
physics-informed networks in the geometry that the energy induces on the parameters
\cite{MullerZeinhofer2023}.

\subsection{Training without solutions, and its comparison with supervision}
Surrogates trained on a physical objective instead of solutions have a history in
computational physics: on a variational objective for elliptic problems with random inputs \cite{Zhu2019},
on the residual of autoregressive dynamics \cite{GenevaZabaras2020} and on irregular domains mapped to a reference grid \cite{Gao2021PhyGeoNet}. Operators have been
trained the same way
\cite{Li2024PINO,WangWangPerdikaris2021}, among them a physics-informed neural operator for
solid mechanics trained, for static problems, on a weak form based on the principle of least
work \cite{Kaewnuratchadasorn2024PINOS}. Variational operator learning \cite{Xu2024VOL}
trains operators on the finite-element system, evaluated matrix-free, and uses the gradient
identity of \cref{lem:exact} as a computational device; its loss is the Euclidean norm of
the residual, or a regression on provisional labels produced by a few conjugate-gradient
iterations. The variational physics-informed neural operator \cite{Eshaghi2025VINO}
evaluates the continuous functional analytically on its own bilinear discretisation over
uniform grids. Larson et~al.~\cite{Larson2026MLPEnergy} train multilayer perceptrons that
map the data of parametrised, possibly nonlinear, problems to the nodal values of their
finite-element solutions by minimising the energy assembled element by element, without precomputed solutions. Do
et~al.~\cite{Do2026MNES} parametrise the nodal displacement
vector of a finite-element mesh by a coordinate-based network and minimise the total
potential energy of the discretised system, for linear and nonlinear plane elasticity.
Other label-free approaches build on the finite-element residual or
weak form \cite{Lee2025FEONet,Uriarte2022,Gao2022,Goswami2022}, and a graph network for the
elastic analysis of braced steel frames was trained without labels on the residual of the
equilibrium between internal and external forces \cite{Song2023StructGNN}. Closest to the present work
are graph networks on finite-element meshes trained by minimising the potential energy
\cite{Dalton2023}, physics-informed MeshGraphNets trained without data for non-stationary
and nonlinear problems \cite{Wurth2024}, a three-dimensional convolutional network for
elastic fields in periodic polycrystals, trained without reference solutions on the
divergence of its predicted stress \cite{MonteiroFernandes2025}, and the pretrain finite element method \cite{Wang2026PretrainFEM}. The last trains a
point-cloud transformer operator on the potential energy, with derivatives from finite-element shape functions, and uses its
prediction to warm-start the finite-element solver; a follow-up applies that approach to
three-dimensional pressure vessels \cite{Sun2026FEVessel}. Several of these works compare
label-free with data-driven training. Physics-constrained surrogates trained on a
variational objective gave predictions comparable to data-driven ones \cite{Zhu2019}. Three
train the same network on the same data with and without reference solutions, in single
runs. In soft-tissue beams the two had comparable displacement errors and the label-free one
smaller errors in energy and deformation gradient, a gap that adding the deformation
gradient to the supervised loss narrowed but did not close \cite{Dalton2023}. In polycrystals the label-free network was more
accurate in stress than the supervised one, which was trained on the reference stresses
\cite{MonteiroFernandes2025}. In pressure vessels the supervised one was more accurate in displacement and in the $H^1$ seminorm
\cite{Sun2026FEVessel} (\cref{sec:discussion:earlier}). Adding the deformation gradient, as in the
soft-tissue beams, is one way of changing the norm of a supervised loss while keeping its
labels; others are reviewed in \cref{sec:related:norm}. We are not aware of a comparison
that trains an elasticity surrogate on its labels in the energy norm of the assembled
stiffness matrix, nor of a comparison with and without labels repeated over several seeds
under criteria fixed in advance.

\subsection{The norm of the loss}\label{sec:related:norm}
Supervised mesh surrogates, from message-passing networks \cite{Pfaff2021,Dalton2022} to
operator networks \cite{Lu2021DeepONet,Li2021FNO} and geometry-aware operators
\cite{He2024GeomDeepONet,Zeng2025PCNO,Liu2026GINOT}, are typically trained and evaluated
with relative $L^2$ errors of the predicted fields, computed on the nodes or points as
Euclidean norms. That the norm of a loss shapes what a
network learns is known in machine learning. Training on derivatives as well as values, Sobolev training, improves the accuracy of the derivatives \cite{Czarnecki2017}, and for
physics-informed networks the $L^2$ residual loss can fail to control the solution where
another norm does \cite{WangL2NeurIPS2022}. For surrogates of partial differential
equations, losses in other norms have been compared with plain ones on the same data. An $H^1$ loss
lowered both the relative $L^2$ and the relative $H^1$ errors of three neural
operators on Darcy flow and on a multiscale elliptic problem \cite{Liu2024Multiscale}, and
adding a stress term, computed by passing the predicted strains through the material model,
lowered the stress errors of a graph network for elastoplastic microstructures
\cite{Storm2024MicroGNN}. Energy-norm losses have been compared with others on the same labels for scalar problems. One surrogate for multiscale Darcy flow predicts the
coefficients of a localised orthogonal decomposition; trained on the energy norm of their
error, it reached lower relative $L^2$ errors of the solution than when trained with $L^2$, relative $L^2$ or $L^1$ losses \cite{Haltmayer2026LOD}. A neural preconditioner for flexible
conjugate gradients trained in the energy norm needed fewer iterations than one trained in
the Euclidean norm \cite{Rudikov2024FCGNO}. For finite-element surrogates of linear
elasticity the right norm is fixed by the problem itself: it is the energy norm, which the
energy supplies without solutions. \Cref{rem:rank} and \cref{sec:checks:ranking} show
that the displacement error can rank surrogates in the opposite order to their stress
errors.

\subsection{Error estimation without the solution}
A posteriori error estimation bounds the energy-norm error of a finite-element approximation
without the exact solution \cite{AinsworthOden2000}: from a statically admissible stress
field through the hypercircle of Prager and Synge \cite{PragerSynge1947}, or through the error in the constitutive relation \cite{LadevezeLeguillon1983}. The latter has
been applied to the solutions of physics-informed networks in elasticity, with the admissible pair of
displacement and stress given by a network of mixed form \cite{GuoHaghighat2022}. Krylov
methods estimate the energy-norm error of an iterate by Gauss quadrature
\cite{GolubMeurant1997,StrakosTichy2002}. The checks of
\cref{sec:theory:metrics,sec:theory:detector} are weaker: they rank predictions and detect gross failures, but need nothing beyond the
assembled stiffness matrix and load vectors. The nearest uses we found are a residual-norm test that decides
whether a warm-started solve is needed \cite{Wang2026PretrainFEM} and an energetic upper
bound on the effective stiffness computed from predicted fields
\cite{MonteiroFernandes2025}.

\subsection{Solvers and reduced bases}
Neural predictions have been used as initial iterates of iterative solvers
\cite{Wang2026PretrainFEM,Eshaghi2026NOWS,Sun2026FEVessel} (supplementary
Section~\SuppWarm{}), inside hybrid iterative schemes \cite{Lee2025MetaSolver}, and to enrich
finite-element approximation spaces \cite{Barucq2026Enrich}. Reduced basis methods, the
classical surrogates of parametrised linear problems \cite{Hesthaven2016RB}, work in one
truth space, usually on a reference domain to which each geometry is mapped, and rely for
efficiency on an affine, or affinely approximated, dependence on the parameters. The instances here differ in the number of holes or cavities, which no such mapping covers, and
in their meshes. The Ritz projection of \cref{sec:methods:ritz} is a classical approximation
that does apply to them, at the price of the stiffness matrix of each instance.
\section{Energy, norm and stress}\label{sec:theory}

This section states what three common losses regress on the assembled system, why the energy
norm is the norm of the stress error, how it relates to the displacement error, and what the
energy of a prediction reveals without a reference solution. \Cref{lem:exact} and
\cref{rem:norms} are restated, with proofs, from the companion note \cite{CaoSong2026}, as
are two statements on conditioning and on gradient descent that supplementary
Section~\SuppRepro{} gives with their checks; the other statements are added here. All are
elementary consequences of the algebra of symmetric positive definite systems
\cite{Ciarlet1978,Saad2003}. What they provide is an exact account of what each loss
optimises, of how the results should be measured, and of what can be checked without
solving. Each statement with numerical content is checked numerically on assembled instances
(supplementary Section~\SuppRepro).

\subsection{Discrete setting}\label{sec:theory:setting}

Consider a linear elastostatic problem on a domain $\Omega\subset\R^{d}$, $d\in\{2,3\}$,
discretised by conforming finite elements, with homogeneous Dirichlet conditions on part
of the boundary. After the constrained degrees of freedom are eliminated, the $n$ free
nodal displacements of the finite-element solution $\Ustar\in\R^{n}$ satisfy
\begin{equation}
K\Ustar = F,
\label{eq:system}
\end{equation}
where $K\in\R^{n\times n}$ is the stiffness matrix restricted to the free degrees of
freedom and $F\in\R^{n}$, $F\neq0$, the load vector. We call $\Ustar$ the reference
solution; with sufficient Dirichlet constraints, $K$ is symmetric positive definite. The
discrete potential energy of a vector $u\in\R^{n}$ is
\begin{equation}
\Pi_h(u) = \tfrac12\,u^{\top}Ku - F^{\top}u ,
\label{eq:energy}
\end{equation}
and the energy norm, which is called the $K$-norm in numerical linear algebra, is
$\norm{v}_K=(v^{\top}Kv)^{1/2}$ for $v\in\R^{n}$. For a prediction $u$ with error
$e=u-\Ustar$ we call $\gap(u)=\Pi_h(u)-\Pi_h(\Ustar)$ its energy gap and
\begin{equation}
g(u) = \frac{\norm{u-\Ustar}_K^{2}}{\norm{\Ustar}_K^{2}}
\label{eq:relgap}
\end{equation}
its relative energy gap. Substituting \eqref{eq:system} into \eqref{eq:energy} gives
$\Pi_h(\Ustar)=-\norm{\Ustar}_K^2/2$, so the zero field has $g(0)=1$. The smallest and
largest eigenvalues of $K$ are $\lambda_{\min}$ and $\lambda_{\max}$, and
$\kappa=\lambda_{\max}/\lambda_{\min}$ is its spectral condition number. Evaluating
$\Pi_h(u)$ requires $K$ and $F$, which a finite-element code assembles before solving,
and one sparse matrix-vector product; it does not require $\Ustar$. The energy gap and $g$
do require $\Ustar$: in this paper they are evaluation quantities, computed on evaluation
instances only. Differences of energy do not (\cref{cor:rank}).

\subsection{Training objective}\label{sec:theory:exact}

\begin{lemma}[Exactness]\label{lem:exact}
For every $u\in\R^{n}$,
\begin{equation}
\Pi_h(u)-\Pi_h(\Ustar) = \tfrac12\norm{u-\Ustar}_K^{2}
\qquad\text{and}\qquad
\nabla\Pi_h(u) = K(u-\Ustar).
\label{eq:exact}
\end{equation}
\end{lemma}

\begin{proof}
Expanding $\norm{u-\Ustar}_K^2/2=u^{\top}Ku/2-u^{\top}K\Ustar
+\Ustar{}^{\top}K\Ustar/2$ and using $K\Ustar=F$ gives
$\norm{u-\Ustar}_K^2/2=\Pi_h(u)+\norm{\Ustar}_K^2/2=\Pi_h(u)-\Pi_h(\Ustar)$.
Differentiating \eqref{eq:energy} gives $\nabla\Pi_h(u)=Ku-F=K(u-\Ustar)$.
\end{proof}

Hence $g(u)=\gap(u)/\lvert\Pi_h(\Ustar)\rvert$: the relative energy gap is the ratio of the
strain energy of the error to that of the solution. A surrogate is trained on many
instances. Let instance $i\in\{1,\dots,N\}$ have stiffness matrix $K_i$ and $J$ load
vectors $F_{ij}$, $j\in\{1,\dots,J\}$, with solutions $\Ustar_{ij}=K_i^{-1}F_{ij}$ and
energies $\Pi_{ij}(u)=u^{\top}K_iu/2-F_{ij}^{\top}u$, and let $u_{ij}(\theta)$ be the
prediction of a network with parameters $\theta$.

\begin{corollary}[Training objective]\label{cor:training}
The label-free objective satisfies
\begin{equation}
\mathcal{L}_{\Pi}(\theta) =
\frac{1}{NJ}\sum_{i=1}^{N}\sum_{j=1}^{J}\Pi_{ij}\bigl(u_{ij}(\theta)\bigr) =
\frac{1}{NJ}\sum_{i=1}^{N}\sum_{j=1}^{J}\tfrac12\norm{u_{ij}(\theta)-\Ustar_{ij}}_{K_i}^{2}
+ \Pi^{\ast},
\label{eq:training}
\end{equation}
where $\Pi^{\ast}=(NJ)^{-1}\sum_{i,j}\Pi_{ij}(\Ustar_{ij})$ does not depend on $\theta$.
Minimising $\mathcal{L}_\Pi$ is therefore supervised regression in the energy norm: the
two objectives have the same minimisers and the same gradient with respect to $\theta$ at
every $\theta$, yet $\mathcal{L}_\Pi$ is evaluated without any $\Ustar_{ij}$.
\end{corollary}

\begin{proof}
Apply \cref{lem:exact} to each term; the chain rule carries the equality of gradients
from $u_{ij}$ to $\theta$.
\end{proof}

The regression in \eqref{eq:training} sums squared energy-norm errors without
normalising them, so instances whose solutions carry more strain energy weigh more.
Normalising each term by $\lvert\Pi_{ij}(\Ustar_{ij})\rvert$ would require the solutions,
so the label-free objective cannot be normalised in this way.

\subsection{Three losses, three norms}\label{sec:theory:norms}

\begin{remark}[Three norms]\label{rem:norms}
With the residual $r(u)=F-Ku=-Ke$, \cref{lem:exact} reads
$\gap(u)=r(u)^{\top}K^{-1}r(u)/2$: the energy gap is half the squared residual
measured in the norm induced by $K^{-1}$, the norm in which a small residual means an
accurate solution. The inverse cancels algebraically in $\Pi_h$, so the objective needs
neither a solve nor a preconditioner. A Euclidean residual loss, by contrast, is
$\norm{r(u)}_2^2=e^{\top}K^{2}e$, a regression in the norm of $K^2$, whose condition number
is $\kappa^2$; the conditioning of residual losses has been linked to the difficulty of
training physics-informed networks \cite{WangYuPerdikaris2022,DeRyck2024}. A displacement
loss is a regression in the Euclidean norm $e^{\top}e$. Let $K=\sum_{m}\lambda_m
w_mw_m^{\top}$ be the eigendecomposition of $K$, with orthonormal eigenvectors $w_m$, and
$\hat e_m=w_m^{\top}e$ the error's component in eigenmode $m$. The three losses weight it by
$1$, $\lambda_m$ and $\lambda_m^2$:
\begin{equation}
\norm{e}_2^2=\sum_m\hat e_m^2,\qquad
\norm{e}_K^2=\sum_m\lambda_m\hat e_m^2,\qquad
\norm{r(u)}_2^2=\sum_m\lambda_m^2\hat e_m^2 .
\label{eq:norms}
\end{equation}
\end{remark}

In elasticity the smooth, low-frequency modes have small eigenvalues and the oscillatory,
high-frequency modes large ones, so an error made of low-frequency modes is large in
displacement and small in energy, and one made of high-frequency modes the reverse. The
displacement loss thus under-weights, relative to the energy, the stiff modes that carry the
strains, and the Euclidean residual over-weights them. Which of these norms measures the
stress is the subject of \cref{prop:stress}.

\subsection{The energy norm and the stress}\label{sec:theory:stress}

Let $v_h$ denote the finite-element displacement field of a vector \mbox{$v\in\R^n$}, extended by
zero on the constrained degrees of freedom, $\varepsilon(v_h)$ its strain,
\mbox{$\sigma(v_h)=\mathbb{C}:\varepsilon(v_h)$} its stress for the fourth-order elasticity
tensor $\mathbb{C}$, $s=\sigma-(\operatorname{tr}\sigma)I/3$ the deviatoric stress, with
$I$ the identity tensor, and
$\vm(\sigma)=\bigl((3/2)\,s:s\bigr)^{1/2}$ the von Mises stress of a stress tensor
$\sigma$; we write $\vm(v_h)$ and $\operatorname{tr}\sigma(v_h)$ for the fields
$\vm(\sigma(v_h))$ and $\operatorname{tr}(\sigma(v_h))$. For a homogeneous isotropic material with
Young's modulus $E$ and Poisson's ratio $\nu$, $G=E/(2(1+\nu))$ is the shear modulus and
$B=E/(3(1-2\nu))$ the bulk modulus.

\begin{proposition}[Energy gap and stress error]\label{prop:stress}
Let the material be homogeneous and isotropic and the stiffness matrix be integrated
exactly, as it is for linear elements. For every $v\in\R^{n}$,
\begin{equation}
\norm{v}_K^{2} = \int_{\Omega}\sigma(v_h):\mathbb{C}^{-1}:\sigma(v_h)\,\mathrm{d}\Omega
= \int_{\Omega}\Bigl(\frac{\vm(v_h)^{2}}{3G}
+\frac{(\operatorname{tr}\sigma(v_h))^{2}}{9B}\Bigr)\mathrm{d}\Omega .
\label{eq:stressnorm}
\end{equation}
Consequently, for a prediction $u$ with error $e=u-\Ustar$,
\begin{equation}
\gap(u) = \frac12\int_{\Omega}\bigl(\sigma(u_h)-\sigma(\Ustar_h)\bigr):\mathbb{C}^{-1}:
\bigl(\sigma(u_h)-\sigma(\Ustar_h)\bigr)\,\mathrm{d}\Omega ,
\label{eq:gapstress}
\end{equation}
and the von Mises fields of $u$ and $\Ustar$ satisfy
\begin{gather}
\bigl\lVert\vm(u_h)-\vm(\Ustar_h)\bigr\rVert_{L^2(\Omega)}^{2}\le 3G\,\norm{u-\Ustar}_K^{2},
\label{eq:vmbound}\\
\frac{\lVert\vm(u_h)-\vm(\Ustar_h)\rVert_{L^2(\Omega)}^{2}}{\lVert\vm(\Ustar_h)\rVert_{L^2(\Omega)}^{2}}
\le (1+\gamma^{\ast})\,g(u),
\label{eq:vmboundrel}
\end{gather}
where $\gamma^{\ast}=G\lVert \operatorname{tr}\sigma(\Ustar_h)\rVert_{L^2(\Omega)}^2/
(3B\lVert\vm(\Ustar_h)\rVert_{L^2(\Omega)}^2)$, defined when $\vm(\Ustar_h)\neq0$, is the
ratio of the volumetric to the deviatoric strain energy of the solution. In plane stress
the same statements hold for the plane stress state, with the three-dimensional moduli $G$
and $B$.
\end{proposition}

\begin{proof}
For a conforming discretisation with exactly integrated stiffness,
\begin{equation}
v^{\top}Kv=\int_\Omega\varepsilon(v_h):\mathbb{C}:\varepsilon(v_h)\,\mathrm{d}\Omega
=\int_\Omega\sigma(v_h):\mathbb{C}^{-1}:\sigma(v_h)\,\mathrm{d}\Omega .
\label{eq:kcompliance}
\end{equation}
For an isotropic
material $\mathbb{C}^{-1}:\sigma=s/(2G)+\operatorname{tr}\sigma\,I/(9B)$, so
$\sigma:\mathbb{C}^{-1}:\sigma=s:s/(2G)+(\operatorname{tr}\sigma)^2/(9B)
=\vm^2/(3G)+(\operatorname{tr}\sigma)^2/(9B)$, which is \eqref{eq:stressnorm}. Equation \eqref{eq:gapstress}
follows from \cref{lem:exact} with $v=e$ and the linearity of $\sigma$. As a function of
the stress tensor, $\vm$ is $(3/2)^{1/2}$ times the Frobenius norm of the deviatoric part,
a seminorm, so that pointwise
$\lvert\vm(u_h)-\vm(\Ustar_h)\rvert\le\vm(\sigma(u_h)-\sigma(\Ustar_h))=\vm(e_h)$;
squaring, integrating and dropping the volumetric term of \eqref{eq:stressnorm} with $v=e$
gives \eqref{eq:vmbound}. With $v=\Ustar$, \eqref{eq:stressnorm} gives
$\norm{\Ustar}_K^2=(1+\gamma^\ast)\lVert\vm(\Ustar_h)\rVert_{L^2}^2/(3G)$, and
\eqref{eq:vmboundrel} follows. In plane stress, the energy density of the plane stress
state equals $\sigma:\mathbb{C}^{-1}:\sigma/2$ with the three-dimensional compliance,
because the out-of-plane strain does no work, and the von Mises stress is the
three-dimensional formula applied to that state.
\end{proof}

Equation \eqref{eq:gapstress} is the sense in which a network trained in the energy norm
learns from a weighted mean-square stress error: by \cref{cor:training}, $\mathcal{L}_\Pi$
is, up to a constant, the mean over the training set of half the integral of the
compliance-weighted squared stress error. The bounds \eqref{eq:vmbound} and
\eqref{eq:vmboundrel} concern the area- or volume-weighted von Mises error, whereas the
metric reported in \cref{sec:methods:metrics} weights elements equally, as pre-registered;
on graded meshes the two differ, so the bounds explain rather than bound the reported
numbers. Measured after the runs for the networks trained on 1,024 instances, the two
differed little (supplementary Section~\SuppRepro).

\subsection{Displacement and energy-norm errors}\label{sec:theory:metrics}

\begin{remark}[Size and spectral position of an error]\label{rem:factor}
For $v\neq0$ let $R(v)=v^{\top}Kv/v^{\top}v$ be the Rayleigh quotient of $K$, so that
$\lambda_{\min}\le R(v)\le\lambda_{\max}$ and, in the eigenbasis of \cref{rem:norms},
$R(e)=\sum_m\lambda_m\hat e_m^2/\sum_m\hat e_m^2$ is the mean eigenvalue of $K$ weighted by
the squared modal coefficients of $e$. For a prediction $u\neq\Ustar$, dividing and
multiplying \eqref{eq:relgap} by $\norm{u-\Ustar}_2^2$ and by $\norm{\Ustar}_2^2$ gives
\begin{equation}
g(u)=\rho(u)\,\delta(u)^{2},
\qquad
\rho(u)=\frac{R(u-\Ustar)}{R(\Ustar)},
\qquad
\delta(u)=\frac{\norm{u-\Ustar}_2}{\norm{\Ustar}_2}:
\label{eq:factor}
\end{equation}
the relative energy gap is the squared relative displacement error $\delta^2$ times the
normalised Rayleigh quotient $\rho$, which measures how far up the spectrum of $K$ the error
lies compared with the solution. For two predictions $u_a$ and $u_b$ of the same load case,
$g(u_a)/g(u_b)=(\rho(u_a)/\rho(u_b))\,(\delta(u_a)/\delta(u_b))^2$, so that a difference
in relative energy gap is a difference in the size of the errors, in their spectral
position,
or in both.
\end{remark}

\begin{remark}[Opposite rankings]\label{rem:rank}
For two predictions with errors $e_a,e_b\neq0$,
\begin{equation}
\kappa^{-1/2} \le \frac{\norm{e_a}_K/\norm{e_b}_K}{\norm{e_a}_2/\norm{e_b}_2} \le \kappa^{1/2},
\label{eq:rank}
\end{equation}
by $\lambda_{\min}\norm{e}_2^2\le\norm{e}_K^2\le\lambda_{\max}\norm{e}_2^2$, that is,
$\kappa^{-1}\le R(e_a)/R(e_b)\le\kappa$ in the notation of \cref{rem:factor}, and both
bounds are attained by eigenvectors of $\lambda_{\max}$ and $\lambda_{\min}$. Hence two
predictions can be ranked in opposite orders by the displacement error and by the
energy-norm error only if their displacement errors differ by less than a factor
$\kappa^{1/2}$, and for every ratio within that factor some pair of errors is so ranked;
\cref{sec:methods:data} gives the range of $\kappa^{1/2}$ on these instances. By
\cref{prop:stress} the energy-norm error is the compliance-weighted stress error, so the
same holds for it; for the von Mises error, \eqref{eq:vmbound} gives a one-sided bound only.
\Cref{sec:checks:ranking} reports such reversals between two trained networks.
\end{remark}

\begin{corollary}[Ranking by energy]\label{cor:rank}
For any two predictions $u_a,u_b$ of the same load case,
$\Pi_h(u_a)-\Pi_h(u_b)=\gap(u_a)-\gap(u_b)
=\bigl(\norm{u_a-\Ustar}_K^2-\norm{u_b-\Ustar}_K^2\bigr)/2$, so the energy ranks
any set of predictions of one load case exactly by their energy-norm errors. Over a
fixed set of load cases, the mean energy of a surrogate's predictions differs from the mean
of their energy gaps by a constant, the mean energy of the solutions, so the mean energy
ranks surrogates by their mean energy gap. Neither statement needs a reference solution.
\end{corollary}

\begin{proof}
Apply \cref{lem:exact} to each prediction and subtract; for the second statement, average
over the load cases.
\end{proof}

Ranking candidate fields by their energy is the Ritz minimum principle itself. The ranking
is by the energy gap, not by the relative energy gap $g$ of the tables, whose normalisation
by $\lvert\Pi_h(\Ustar)\rvert$ needs the solution; the tables report $g$ because it puts
instances and load cases of different strain energy on one scale.

\subsection{Checks without the solution}\label{sec:theory:detector}

\begin{corollary}[Zero-field test]\label{cor:zero}
For each load case, $\Pi_h(u)>0$ if and only if $\norm{u-\Ustar}_K>\norm{\Ustar}_K$, that
is, if and only if $g(u)>1$: the prediction is further from the solution in the energy
norm than the zero field. The test needs no reference solution.
\end{corollary}

\begin{proof}
Apply \cref{lem:exact} at $u$ and at $0$, and use $\Pi_h(0)=0$.
\end{proof}

What \cref{cor:zero} adds to the Ritz principle for surrogates is a test against a fixed
reference, the zero field, that each prediction passes or fails on its own. It detects
gross failures only: it does not bound the error of a prediction that passes it. The counts
of predictions worse than the zero field in the tables use the mean of $g$ over the four
load cases of an instance, as the evaluation records it, and so need the solutions. The counts of single load cases in
\cref{sec:checks:zero} coincide exactly with the test of
\cref{cor:zero}.

\begin{proposition}[Energy-optimal amplitude]\label{prop:amp}
Let $u\in\R^{n}$ with $u^{\top}Ku>0$ and let
\begin{equation}
\cstar = \frac{F^{\top}u}{u^{\top}Ku}.
\label{eq:cstar}
\end{equation}
Then $\cstar$ minimises $\Pi_h(cu)$ over $c\in\R$,
\begin{equation}
\Pi_h(\cstar u) = -\frac{(F^{\top}u)^{2}}{2\,u^{\top}Ku} \le \min\{\Pi_h(u),\,0\},
\qquad
g(\cstar u) = 1-\cos^{2}\psi ,
\label{eq:amp}
\end{equation}
where $\cos\psi=u^{\top}K\Ustar/(\norm{u}_K\norm{\Ustar}_K)$; in particular
$g(\cstar u)\le\min\{g(u),1\}$.
\end{proposition}

\begin{proof}
$\Pi_h(cu)=c^2u^{\top}Ku/2-cF^{\top}u$ is a convex quadratic in $c$ with minimiser
$\cstar$ and minimum $-(F^{\top}u)^2/(2u^{\top}Ku)$; since $c=1$ and $c=0$ are admissible,
the minimum is at most $\Pi_h(u)$ and $\Pi_h(0)=0$. By \cref{lem:exact} and
$\Pi_h(\Ustar)=-\norm{\Ustar}_K^2/2$, $g(v)=1+2\Pi_h(v)/\norm{\Ustar}_K^2$ for every
$v\in\R^{n}$; with $F^{\top}u=\Ustar{}^{\top}Ku$ this gives $g(\cstar u)=1-\cos^2\psi$.
\end{proof}

The rescaled prediction $\cstar u$ is the $K$-orthogonal projection of $\Ustar$ onto the
span of $u$, the one-dimensional Ritz approximation. It needs no label and no work beyond
evaluating $\Pi_h(u)$, since $u^{\top}Ku$ and $F^{\top}u$ are the two terms of $\Pi_h(u)$.
After the rescaling only the error of direction, $\sin^2\psi$, remains; an error of
amplitude is removed exactly. The improvement is in the energy norm: in other norms,
such as the displacement error, rescaling can make a prediction worse. Together,
\cref{cor:zero} and \cref{prop:amp} give a deployment guard: compute $\Pi_h(u)$, flag the
prediction if it is positive, and rescale it by $\cstar$, which never makes it worse in the
energy norm and makes it no worse than the zero field.
\section{Problems, networks and comparisons}\label{sec:methods}

\subsection{Problems and data}\label{sec:methods:data}

Two families of linear elastostatic problems were used, one in two dimensions and one in
three. Both are clamped on the face $x=0$ and carry four load cases that share one stiffness
matrix: a downward and an axial traction on the face opposite the clamp, a shear traction on
the top face, and a downward body force. The magnitude of each load case is drawn
independently as $0.05$ times a number uniform on $[0.5,1.5]$. Young's modulus is 1 and
Poisson's ratio is uniform on $[0.25,0.38]$ per instance; all quantities are dimensionless.
The problems are linear, so the fields scale with the loads and inversely with Young's
modulus; the values in the figures can be read in any consistent units, in which a physical
load makes the strains small.
Meshes were generated with Gmsh \cite{Geuzaine2009}, and reference solutions were computed
by sparse direct factorisation.

\subsubsection{Two dimensions}
Rectangular plates $[0,l_x]\times[0,l_y]$ in plane stress, with $l_x$ uniform on $[1.5,3]$,
$l_y$ uniform on $[0.8,1.5]$ and zero to three non-overlapping circular holes of radius
between $0.06$ and $0.16$ times $\min(l_x,l_y)$, are meshed with linear triangles of target
size uniform on $[0.05,0.12]$ and assembled with scikit-fem \cite{Gustafsson2020}. The
corpus has 30,000 instances: 256 form the validation set and the first 1,024 of the
remainder the training pool. The last \numHoldTwoTail{} instances of the remainder lie
beyond the deepest prefix of it that any recorded run read, and \numHoldTwoN{} of them form
the holdout set of \cref{sec:holdout}.

\subsubsection{Three dimensions}
Boxes $[0,l_x]\times[0,l_y]\times[0,l_z]$, with $l_x$ uniform on $[1.5,3]$, $l_y$ on
$[0.8,1.5]$, $l_z$ on $[0.6,1.2]$ and zero to three non-overlapping spherical cavities of
radius between $0.08$ and $0.16$ times $\min(l_x,l_y,l_z)$, are meshed with linear
tetrahedra at a characteristic length $l_{\mathrm{c}}$ drawn uniformly from $[0.0579,0.0906]$ per
instance, the in-band range. The corpus has 2,000 instances: 256 form the validation set, the first 1,024 of
the remainder the training pool, and the other \numHoldThreeTail{} were not used for
training or validation; \numHoldThreeN{} of them form the holdout set of \cref{sec:holdout}.
The validation meshes have between \numNodesInMin{} and \numNodesInMax{} nodes (median
\numNodesInMedian{}), with three degrees of freedom per node. A finer set of 320 instances
from the same distribution is meshed at $l_{\mathrm{c}}=0.0374$, about half the element size. Its first 256 instances, with between \numNodesFineMin{} and \numNodesFineMax{} nodes (median
\numNodesFineMedian{}), are used to test transfer to a finer mesh, and the last 64 are reserved for
few-shot training.

\subsubsection{Conditioning}
On sixteen assembled validation instances of each corpus, the spectral condition number
$\kappa$ of the stiffness matrix ranged from \numKappaTwoLow{} to \numKappaTwoHigh{} in two
dimensions and from \numKappaThreeLow{} to \numKappaThreeHigh{} in three, so
$\kappa^{1/2}$ lies between about \numSqrtKappaMin{} and \numSqrtKappaMax{}.

\subsection{Networks}\label{sec:methods:networks}

\subsubsection{Transformer}
Every mesh node is a token. Its input features are the node's coordinates, centred and
scaled to unit root-mean-square radius; its Dirichlet flags; and its consistent nodal load
for the load case, divided by $F_{\max}$, the largest absolute nodal load component over the
instance's four load cases. Two global descriptors are appended to every token. The load summary holds the total load
magnitude and the net load per component, each divided by the number of nodes and by
$F_{\max}$, and the fraction of loaded nodes. The geometry descriptor holds the extents, each
divided by the upper end of its sampling range, so that the absolute size is kept; Poisson's
ratio; the number of holes or cavities and their area or volume fraction; and, in two
dimensions, their mean radius. An encoder of 8 pre-normalised transformer blocks
\cite{Vaswani2017} of width 256 with 8 attention heads applies full self-attention over the nodes. A decoder maps each node's latent vector, concatenated with the mean latent vector
over the nodes, to the node's displacement through a three-layer perceptron, and the
constrained components are set to zero. Each load case is a separate sequence; the four
sequences of an instance are processed as one batch. The encoder and decoder have
\numParamsTransformer{} million parameters. The decoded field is multiplied by $F_{\max}$,
so that the prediction is proportional to the load amplitude; since $F_{\max}$, a nodal
load, shrinks with the element size, this scaling makes the output depend on the mesh
(\cref{sec:discussion:transfer}).

\subsubsection{Graph network}
As a supervised baseline of a widely used kind we used a MeshGraphNets-style
encode-process-decode network \cite{Pfaff2021} with node and edge encoders, 8
message-passing layers of width 256 over the mesh edges (edge features: relative position
and length), the same node features and the same output scaling as the transformer. It has \numParamsGraph{} million parameters.

\subsubsection{Baselines}
The accuracy tables also report three deterministic baselines: the zero field; a
nearest-neighbour field, the solution of the labelled training instance nearest in the
geometry descriptor, interpolated to the new mesh in normalised coordinates; and a scaled
polynomial, a quadratic ridge regression of the field's direction on the node features with
its norm regressed on the instance's descriptors. The last two use the same labelled
instances as the supervised networks.

\subsection{Losses and training}\label{sec:methods:training}

The label-free transformer minimises \eqref{eq:training} with $J=4$: at each step one
training instance is drawn, without replacement within an epoch, its four load cases are
predicted, and the mean of their energies is the loss; no reference solution is read. The
displacement-supervised transformer, the same network, minimises on labelled instances the
mean over the load cases of the relative displacement error,
\begin{equation}
\mathcal{L}_D=\frac{1}{J}\sum_{j=1}^{J}\frac{\norm{u_j-\Ustar_j}_2}{\norm{\Ustar_j}_2},
\label{eq:ld}
\end{equation}
where $u_j$ is the prediction and $\Ustar_j$ the reference solution for load case $j$ of an
instance, and $F_j$ its load vector. The energy-norm-supervised transformer, again the same
network, minimises the relative energy-norm error,
\begin{equation}
\mathcal{L}_K=\frac{1}{J}\sum_{j=1}^{J}\frac{\norm{u_j-\Ustar_j}_K}{\norm{\Ustar_j}_K}
=\frac{1}{J}\sum_{j=1}^{J}g(u_j)^{1/2},
\label{eq:lk}
\end{equation}
which replaces the norm of \eqref{eq:ld} and nothing else. Since
$\norm{u_j-\Ustar_j}_K^2=u_j^{\top}Ku_j-2F_j^{\top}u_j+F_j^{\top}\Ustar_j$, the loss
$\mathcal{L}_K$ depends on each reference solution only through the number
$\norm{\Ustar_j}_K^2=F_j^{\top}\Ustar_j$, and its gradient with respect to $u_j$,
\begin{equation}
\nabla_{u_j}\mathcal{L}_K=\frac{Ku_j-F_j}{J\,\norm{u_j-\Ustar_j}_K\,\norm{\Ustar_j}_K},
\label{eq:lkgrad}
\end{equation}
is that of the label-free objective of one instance, $(Ku_j-F_j)/J$, weighted per load case.
The two energy-norm losses thus differ in the weighting of the load cases and in the labels,
not in the norm. In the code the squared norm is computed from the difference
$u_j-\Ustar_j$, not from the expansion above, which cancels where the error is small. A negative value from rounding is set to zero, and $10^{-30}$ is added under the square root,
so that the gradient stays defined where the error vanishes. In three dimensions a fourth
variant, the displacement-supervised
transformer with an energy term, adds the mean energy of the four load cases to $\mathcal{L}_D$. The gradient of that term is
scaled to at most the norm of the gradient of $\mathcal{L}_D$, except with 16 labelled
instances, where the term has a fixed weight (supplementary Section~\SuppPrereg). The graph
network is trained on $\mathcal{L}_D$.

All networks were trained with AdamW \cite{Loshchilov2019}, with weight decay $10^{-4}$ and
the gradient norm clipped at 1, for 200 epochs. After a linear warm-up over the first 5\% of the steps to a peak of $10^{-3}$ for the
label-free and $1.5\times10^{-3}$ for the supervised networks, the learning rate followed a
cosine schedule. These rates were fixed before
the runs and not tuned on them, and the energy-norm-supervised transformer used the
displacement-supervised settings unchanged. Training was in single precision with
TensorFloat-32 (TF32) matrix products, with three seeds (0, 1, 2) per network. The networks
are implemented in PyTorch, and the report of each run records the versions of Python,
PyTorch, NumPy and SciPy with which it ran (supplementary Section~\SuppRepro{}). The
supervised networks were trained with 16, 64, 256 and 1,024 labelled instances, the first 16, 64, 256 and 1,024 instances of the training pool. The graph network was trained
with 64 and 1,024 only, and the three-dimensional energy-norm-supervised transformer with
1,024 only, to save machine time, as the pre-registrations of September and October 2026
fixed.
Every network ran 200 epochs, so a supervised network with $N$ labelled instances took
$200N$ steps, 3,200 with 16, and the comparisons at the smaller budgets differ in the number
of steps as well as in the labels. The label-free transformer was trained on the same 1,024
instances without their solutions, for 204,800 steps; its training set is thus the
supervised networks' largest one with the labels removed. Each network is evaluated as it
stands at the end of its last epoch; no checkpoint was selected on the validation instances.

The two norms were compared with everything else held fixed. In the two-dimensional
comparison, at a given seed the transformers start from the same initial weights and, with
1,024 instances, see them in the same sequence of per-epoch permutations, so that the
displacement-supervised and the energy-norm-supervised transformers differ only in the loss.
The three-dimensional energy-norm-supervised transformer was trained in October 2026, about
seven weeks after the displacement-supervised transformer with which it is compared. At each seed it started from that network's initial weights and saw
the instances in its order, as a regression test of the two versions of the code on small
problems showed. The two-dimensional
label-free transformers were trained, with these settings, in a separate pre-registered run
and reused without retraining (supplementary Section~\SuppSide).

\subsection{Metrics}\label{sec:methods:metrics}

For each validation instance and load case we compute five metrics. The relative
displacement error is $\norm{u-\Ustar}_2/\norm{\Ustar}_2$. The relative energy gap is $g(u)$
of \eqref{eq:relgap}, computed as $\gap(u)/\lvert\Pi_h(\Ustar)\rvert$; its square root
$g^{1/2}$ is the relative energy-norm error. With the element-wise von Mises stresses
$\vm(u)$ and $\vm(\Ustar)$, which are constant on linear elements, the von Mises error is
$\norm{\vm(u)-\vm(\Ustar)}_2/\norm{\vm(\Ustar)}_2$, the norms taken over the vector of element values. The peak stress error is
$\lvert\max\vm(u)-\max\vm(\Ustar)\rvert/\max\vm(\Ustar)$, and the critical-region recall is
the share of the elements with the largest $\vm(\Ustar)$, a tenth of all the elements
rounded to the nearest integer, that are also among as many elements with the largest
$\vm(u)$. Every error is relative to the finite-element solution on the same mesh. Where
the clamped boundary meets the rest of the boundary, the stresses of the continuous problem
are in general singular, and the largest element stress there grows as the mesh is refined. Where the peak lies there, the peak stress error measures the agreement with the peak of the
discrete solution, not with a converged value. Each metric is averaged over the four load
cases of an instance and then over
the 256 validation instances; tables give the mean and the sample standard deviation over
the three seeds. The relative energy gap is the primary metric, as pre-registered: by
\cref{prop:stress} it is the strain energy of the stress error relative to that of the
solution. Being a squared error, it changes by the square of the factor by which the
energy-norm error changes; we quote factors of the energy-norm error where the two could be
confused. The normalisation of $g$ needs the solution, so $g$ is computed on evaluation
instances only; without the solution, the energy gives the ranking of \cref{cor:rank} and
the test of \cref{cor:zero}. The tables count as worse than the zero field the instance-seed
predictions whose relative energy gap, averaged over the four load cases, exceeds one.

\subsection{A classical approximation of the same size}\label{sec:methods:ritz}

For each instance we also computed the Ritz-Galerkin projection of the finite-element
problem onto a space of vector polynomials, from the instance's own stiffness matrix and
load vectors, without labels or training. With the instance's box
$[0,l_x]\times[0,l_y]\times[0,l_z]$ and the Legendre polynomials $P_{p_x}$, $P_{p_y}$ and
$P_{p_z}$ of degrees $p_x$, $p_y$ and $p_z$, the trial space of each displacement component
is spanned by the functions
\begin{equation}
\phi(x,y,z)=\frac{x}{l_x}\,P_{p_x}\Bigl(\frac{2x}{l_x}-1\Bigr)\,P_{p_y}\Bigl(\frac{2y}{l_y}-1\Bigr)\,
P_{p_z}\Bigl(\frac{2z}{l_z}-1\Bigr),
\label{eq:ritzbasis}
\end{equation}
for $p_x+p_y+p_z\le p-1$, evaluated at the nodes, and in two dimensions by the same
functions without the factor in $z$; the factor $x/l_x$ makes every function vanish on the
clamped face. With $\Phi$ the matrix whose columns are these functions on the free degrees
of freedom, the projection of a
load case is $u_{\mathrm{Ritz}}=\Phi\,(\Phi^{\top}K\Phi)^{+}\Phi^{\top}F$, the pseudo-inverse taken
through the eigendecomposition of $\Phi^{\top}K\Phi$ with the directions whose eigenvalue is
at most $10^{-13}$ times the largest dropped; on no instance was a direction dropped. Like
the rescaled prediction $\cstar u$, which is the projection onto the span of $u$
(\cref{prop:amp}), $u_{\mathrm{Ritz}}$ minimises $\Pi_h$ over its space. We report every degree of the
ladder that the specification fixed, $p=6$, 7, 8 and 9 in three dimensions, with
\numRitzThreeK{} unknowns, and $p=6$ and 8 in two, with \numRitzTwoK{}, and pair each
instance's projection with every network's prediction of it. This comparison was added after
the pre-registered runs and was not pre-registered. Its specification was written on 20
October 2026, after exploratory readings on instances drawn afresh from the same generators
and on the evaluation sets of a separate study. It was in the public repository before any
paired computation on the instances of this paper, which ran on 28 October 2026. The
projection needs the assembled stiffness matrix, which the surrogates do not.

\subsection{Comparisons fixed in advance}\label{sec:methods:prereg}

Weak baselines and selective reporting have led to over-optimism in machine learning for
fluid-related partial differential equations \cite{McGreivy2024}, and a fair comparison of
two neural operators needed common data and protocols \cite{Lu2022Fair}. Pre-registration guards against these biases in other sciences \cite{Nosek2018}: hypotheses,
analyses and decision criteria are recorded in a dated document before the data are seen.
Four runs decide the claims of this paper, and each was pre-registered. The three-dimensional
comparison (22 to 28 September 2026) trained the label-free transformer and reused the
supervised networks, the graph network among them, that its initial run had trained in the
first half of September 2026. The two-dimensional comparison (8 to 9 October 2026) trained the
displacement-supervised and the energy-norm-supervised transformers and the graph network. The
holdout evaluation (21 October 2026) read the claims of both comparisons again on held-out
instances without training. The fourth is the three-dimensional energy-norm run (22 to 25
October 2026). Each was specified, before it was executed, in a
document that fixes the configuration, the metrics and the criteria, and that declares both outcomes publishable. The documents were committed to the project's public repository before
their runs, and the code started no run whose configuration differed from the registered
hash. Where a comparison of two networks over three seeds carries a verdict, it
uses a noise guard fixed in advance \eqref{eq:guard}: a difference counts only beyond the larger of 10\% and twice its standard error over the seeds. The guard is a rule of
decision, not a test at a stated level of significance, and within it no difference is shown,
which is not equivalence. \Cref{tab:claims} lists the pre-registered claims with their outcomes,
and \cref{sec:holdout} the claims that the holdout evaluation, also pre-registered, read again on held-out instances. Departures from the plans that bear on a result are stated
with it, and the readings taken after the runs, and the Ritz baseline, are marked as such
where they are reported. Supplementary Section~\SuppPrereg{} gives the full record. Every
table, figure and quoted number is computed from the committed records by the repository's
generation script and checked by its tests (supplementary Section~\SuppRepro{}). That script, like the code of the experiments, was written with the help of the generative AI
models Claude Fable 5, Claude Fable 5.1 and Claude Opus 5.5 (Anthropic). Two earlier
pre-registered runs in two dimensions, in July 2026, used an earlier version of the code,
which did not multiply the decoded field by $F_{\max}$. The main one, the July run, is
reported in supplementary Section~\SuppJuly{}; only its test of a warm start
(\cref{sec:cost:labels}) and its transfer readings (\cref{sec:discussion:transfer}) are quoted here.
\section{The norm of the loss}\label{sec:norm}

This section tests the explanation of \cref{sec:theory} directly. The same transformer, with
the same labelled instances, initial weights, order of the instances and learning rate, is
trained with the relative displacement error $\mathcal{L}_D$ \eqref{eq:ld} or with the
relative energy-norm error $\mathcal{L}_K$ \eqref{eq:lk}; nothing else changes
(\cref{sec:methods:training}). Both comparisons were pre-registered, with directions fixed
in advance. \Cref{fig:normlabels} summarises them, and \cref{tab:3d,tab:2d} give every
metric.

\subsection{Three dimensions}\label{sec:norm:three}

\begin{table}[tbp]
\centering
\footnotesize
\setlength{\tabcolsep}{4pt}
\caption{Three-dimensional solids: every metric on the 256 validation instances, four load cases each, for the networks trained on 1,024 instances (the label-free transformer on the same instances without their solutions) and for the baselines. Trained networks: mean $\pm$ sample standard deviation over three seeds. Metrics as in \cref{sec:methods:metrics}; lower is better except in critical-region recall.}
\label{tab:3d}
\resizebox{\ifdim\width>\textwidth\textwidth\else\width\fi}{!}{%
\begin{tabular}{lcccccc}
\toprule
Model & \shortstack{Displacement \\ error} & \shortstack{Relative \\ energy gap} & \shortstack{von Mises \\ error} & \shortstack{Peak stress \\ error} & \shortstack{Critical-region \\ recall} & \shortstack{Worse than \\ zero field} \\
\midrule
Label-free & 0.0300 $\pm$ 0.0028 & 0.00912 $\pm$ 0.00228 & 0.0554 $\pm$ 0.0045 & 0.0807 $\pm$ 0.0134 & 0.793 $\pm$ 0.022 & 0/768 \\
Supervised, displacement & 0.0470 $\pm$ 0.0038 & 31.2 $\pm$ 53.5 & 0.370 $\pm$ 0.140 & 0.784 $\pm$ 0.807 & 0.619 $\pm$ 0.005 & 18/768 \\
Supervised, displacement + energy term & 0.0470 $\pm$ 0.0122 & 0.116 $\pm$ 0.050 & 0.128 $\pm$ 0.008 & 0.115 $\pm$ 0.026 & 0.730 $\pm$ 0.006 & 2/768 \\
Supervised, energy norm & 0.0227 $\pm$ 0.0022 & 0.00504 $\pm$ 0.00022 & 0.0440 $\pm$ 0.0014 & 0.0386 $\pm$ 0.0011 & 0.858 $\pm$ 0.001 & 0/768 \\
Graph network, displacement & 0.0168 $\pm$ 0.0001 & 0.447 $\pm$ 0.084 & 0.296 $\pm$ 0.010 & 0.488 $\pm$ 0.011 & 0.679 $\pm$ 0.009 & 28/768 \\
\midrule
Nearest-neighbour field & 0.382 & 1.22 & 0.717 & 1.23 & 0.509 & 88/256 \\
Scaled polynomial & 1.56 & 4,362 & 14.9 & 54.8 & 0.137 & 256/256 \\
\midrule
Ritz projection, $p = 6$ & 0.0208 & 0.0183 & 0.0854 & 0.321 & 0.811 & 0/256 \\
Ritz projection, $p = 7$ & 0.0134 & 0.0114 & 0.0624 & 0.283 & 0.859 & 0/256 \\
Ritz projection, $p = 8$ & 0.0100 & 0.00838 & 0.0521 & 0.253 & 0.879 & 0/256 \\
Ritz projection, $p = 9$ & 0.00798 & 0.00654 & 0.0449 & 0.223 & 0.896 & 0/256 \\
\midrule
Zero field & 1.00 & 1.00 & 1.00 & 1.00 & -- & -- \\
\bottomrule
\end{tabular}}

\smallskip
\begin{minipage}{\linewidth}\footnotesize\raggedright
The Ritz projection of degree $p$ (\cref{sec:methods:ritz}) was added after the pre-registered runs. This projection, the nearest-neighbour field and the scaled polynomial are deterministic, so their last column counts instances.
\end{minipage}
\end{table}

The three-dimensional energy-norm run trained the transformer on $\mathcal{L}_K$ with the
1,024 labelled instances of the three-dimensional comparison. It also evaluated the stored label-free and
displacement-supervised transformers, which reproduced their earlier values exactly. Its two hypotheses were supported. Averaged over the seeds, the
energy-norm-supervised transformer's median relative energy gap was \numKdKnormGapMedian{},
\numKdGapRel{} lower than the displacement-supervised transformer's \numKdLabGapMedian{} and
beyond the noise guard of \numKdGapTau{}. This is a \numNlThreeNormE{}-fold reduction of the
median energy-norm error. Its median von Mises error was \numKdVmRel{} lower (guard \numKdVmTau{}).
Both medians were lower in each seed. Its displacement error fell too, from
\numThreeLabDisp{} to \numThreeKnormDisp{} in the seed mean (\cref{tab:3d}), and none of its
\numThreePairs{} instance-seed predictions was worse than the zero field, against
\numThreeLabWorse{} of the displacement-supervised transformer's. The verdicts were based on each seed's median over the 256 validation instances. That statistic was
fixed after the per-instance values of the two stored transformers, and the two-dimensional comparison's readings on the
medians, were known. The reason is one validation instance, No.~\numWorstIndex{}: in seed
\numThreeLabWorstSeed{} the displacement-supervised transformer's prediction of it had a
relative energy gap of \numThreeLabWorst{}, averaged over its four load cases. This one value dominates the network's seed means
(\cref{tab:3d}) and widens the guard on them. On the seed means
the von Mises error was \numKdMeansVmRel{} lower, beyond the guard of \numKdMeansVmTau{},
although with three seeds its Welch 95\% interval runs from \numKdMeansVmWelchLow{} to \numKdMeansVmWelchHigh{}. The relative energy gap was
\numKdMeansGapRel{} lower on the seed means, within a guard that this instance widens to \numKdMeansGapTau{}. On the \numHoldThreeN{} held-out
instances of \cref{sec:holdout}, read without a verdict, the two reductions were
\numKdHoldGapRel{} and \numKdHoldVmRel{}, beyond guards of \numKdHoldGapTau{} and
\numKdHoldVmTau{}. The test rests on one model size and one family of geometries, with the
displacement-supervised settings unchanged and not tuned for the new loss. It supports the explanation of \cref{sec:theory} rather than establishing it in general.

\subsection{Two dimensions}\label{sec:norm:two}

\begin{table}[tbp]
\centering
\footnotesize
\setlength{\tabcolsep}{4pt}
\caption{Two-dimensional plates: every metric on the 256 validation instances, four load cases each, for the networks trained on 1,024 instances (the label-free transformer on the same instances without their solutions) and for the baselines. Trained networks: mean $\pm$ sample standard deviation over three seeds. Metrics as in \cref{sec:methods:metrics}; lower is better except in critical-region recall.}
\label{tab:2d}
\resizebox{\ifdim\width>\textwidth\textwidth\else\width\fi}{!}{%
\begin{tabular}{lcccccc}
\toprule
Model & \shortstack{Displacement \\ error} & \shortstack{Relative \\ energy gap} & \shortstack{von Mises \\ error} & \shortstack{Peak stress \\ error} & \shortstack{Critical-region \\ recall} & \shortstack{Worse than \\ zero field} \\
\midrule
Label-free & 0.0634 $\pm$ 0.0031 & 0.0383 $\pm$ 0.0019 & 0.123 $\pm$ 0.002 & 0.128 $\pm$ 0.001 & 0.753 $\pm$ 0.001 & 0/768 \\
Supervised, displacement & 0.0505 $\pm$ 0.0023 & 0.133 $\pm$ 0.014 & 0.235 $\pm$ 0.015 & 0.141 $\pm$ 0.007 & 0.668 $\pm$ 0.015 & 0/768 \\
Supervised, energy norm & 0.0547 $\pm$ 0.0094 & 0.0290 $\pm$ 0.0125 & 0.106 $\pm$ 0.020 & 0.0947 $\pm$ 0.0356 & 0.804 $\pm$ 0.055 & 0/768 \\
Graph network, displacement & 0.0499 $\pm$ 0.0003 & 0.825 $\pm$ 0.020 & 0.578 $\pm$ 0.009 & 0.321 $\pm$ 0.011 & 0.517 $\pm$ 0.006 & 187/768 \\
\midrule
Nearest-neighbour field & 0.404 & 0.959 & 0.735 & 0.643 & 0.515 & 94/256 \\
Scaled polynomial & 1.42 & 873 & 11.4 & 30.8 & 0.159 & 256/256 \\
\midrule
Ritz projection, $p = 6$ & 0.0500 & 0.0404 & 0.122 & 0.272 & 0.828 & 0/256 \\
Ritz projection, $p = 8$ & 0.0424 & 0.0333 & 0.109 & 0.208 & 0.852 & 0/256 \\
\midrule
Zero field & 1.00 & 1.00 & 1.00 & 1.00 & -- & -- \\
\bottomrule
\end{tabular}}

\smallskip
\begin{minipage}{\linewidth}\footnotesize\raggedright
The Ritz projection of degree $p$ (\cref{sec:methods:ritz}) was added after the pre-registered runs. This projection, the nearest-neighbour field and the scaled polynomial are deterministic, so their last column counts instances.
\end{minipage}
\end{table}

The two-dimensional comparison trained the displacement-supervised and the
energy-norm-supervised transformers, which differ only in the loss, and pre-registered two hypotheses about them. H2a holds that the energy-norm-supervised
transformer has the lower relative energy gap, which is close to a check that the change of
loss took effect; H2b holds that it has the lower von Mises error. The von Mises error is a term of neither loss: H2b
tests the explanation that \cref{prop:stress} gives. The comparison was evaluated on the
validation instances on which the results that motivated it had been seen. It was registered after the values of
its label-free transformers on them were known, and after a functional check at a far smaller scale had shown the directions of its hypotheses
(supplementary Section~\SuppPrereg{}). Both were supported on the validation instances. With
1,024 labelled instances the energy-norm-supervised transformer's relative energy gap was
\numHTwoARel{} lower (guard \numHTwoATau{}), \numHTwoARootRel{} lower in the
root-mean-square relative energy-norm error, and its von Mises error \numHTwoBRel{} lower,
\numCmKnormVm{} against \numCmLabVm{} (guard \numHTwoBTau{}). On the seeds' medians the
guard gave the same readings, and the 95\% intervals of the two relative differences, from
Welch's statistic over the seeds and from resampling the instances, lay below zero. Instance
by instance it had the lower relative energy gap on \numPairCmKnormLabGap{} of the
instance-seed pairs, the lower von Mises error on \numPairCmKnormLabVm{} and the lower
displacement error on \numPairCmKnormLabDisp{}: changing the norm lowered the stress error
by more than half and left the displacement error about where it was, \numCmKnormDisp{}
against \numCmLabDisp{}. Weighted by element area, as in \cref{prop:stress}, instead of
equally, the ratio of the two networks' mean von Mises errors was \numSpecVmAreaRatio{},
against \numSpecVmElemRatio{} for the reported metric. On the held-out instances the
reduction of the von Mises error replicated, by more than half, and that of the mean
relative energy gap did not (\cref{sec:holdout}).

\subsection{Where the errors lie in the spectrum}\label{sec:norm:spectrum}

\begin{figure}[tbp]
\centering
\figfull{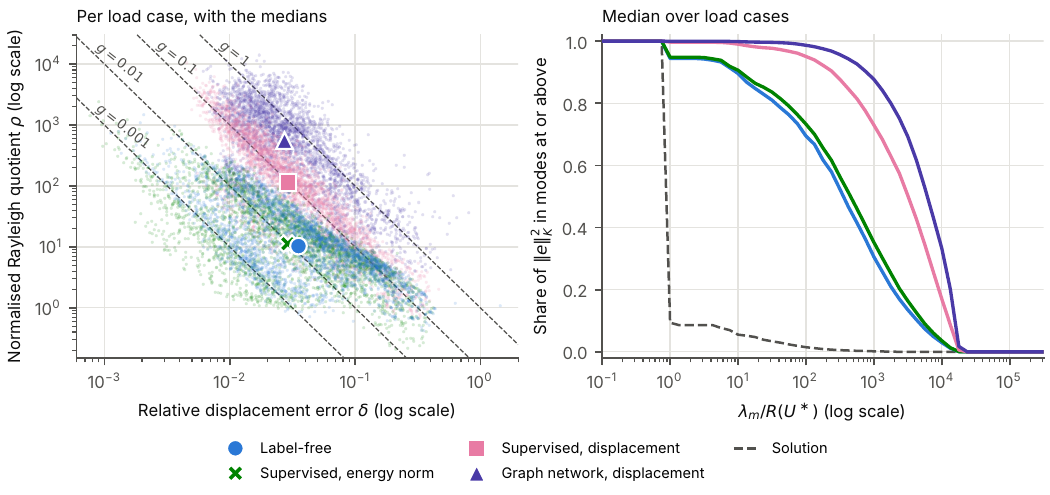}
\caption{Where the errors lie in the spectrum of the stiffness matrix (two dimensions,
networks trained on 1,024 instances; post hoc). Left: normalised Rayleigh quotient $\rho$ of
the error against the relative displacement error $\delta$, one point per seed, validation
instance and load case; large markers: medians; dashed lines: equal relative energy gap
$g=\rho\,\delta^2$. Right: median share of the error's squared energy norm in the
eigenmodes with $\lambda_m/R(\Ustar)$ at or above the abscissa ($\lambda_m$: eigenvalue;
$R(\Ustar)$: the solution's Rayleigh quotient); dashed curve: the solution's share.}
\label{fig:spectra}
\end{figure}

\Cref{rem:factor} writes the relative energy gap of a prediction as the product of the
squared relative displacement error $\delta^2$ and the normalised Rayleigh quotient $\rho$,
which measures how far up the spectrum of $K$ the error lies relative to the solution. After
the two-dimensional run, and without pre-registration, we measured both for the four networks trained on 1,024 instances, on each of the
\numSpecLoads{} triples of seed, validation instance and load case. From the
eigendecomposition of each instance's stiffness matrix we also measured how each error's squared energy norm is distributed over the eigenmodes
(\cref{fig:spectra}). The solution itself lies at the bottom of the spectrum: its Rayleigh
quotient was a median \numSpecRqStarOverMin{} times the smallest eigenvalue of $K$. The
median of $\rho$ was \numSpecRqFree{} for the label-free transformer, \numSpecRqKnorm{} for
the energy-norm-supervised transformer, \numSpecRqLab{} for the displacement-supervised
transformer and \numSpecRqMgn{} for the graph network.

The two transformers that differ only in the norm of the loss had errors of about the same
size in different places. Over the triples, the geometric mean of the ratio of their $\rho$,
displacement-supervised over energy-norm-supervised, was \numSpecLabKnormRq{}
(\numSpecLabKnormRqSeeds{} in seeds 0, 1 and 2), and the displacement-supervised
transformer's $\rho$ was the larger on \numSpecLabKnormAbove{} of the triples. The geometric mean of the ratio of their $\delta^2$ was \numSpecLabKnormDisp{}. On a logarithmic scale the
ratio of $\rho$ accounts for \numSpecSpectralShare{} of the geometric-mean ratio of their
relative energy gaps, \numSpecLabKnormGap{}, and the ratio of $\delta^2$ for the rest. The
modes with $\lambda_m\ge 100\,R(\Ustar)$ held a median \numSpecShareKLab{} of the squared
energy norm of the displacement-supervised transformer's error, against
\numSpecShareKKnorm{} for the energy-norm-supervised transformer, \numSpecShareKFree{} for
the label-free transformer and \numSpecShareKStar{} of the solution's own. The graph
network's errors had the size of the displacement-supervised transformer's (ratio of
$\delta^2$ \numSpecMgnLabDisp{}) and lay higher still (ratio of $\rho$ \numSpecMgnLabRq{}).
The networks ran with TF32 matrix products, whose rounding weighs heavily in the energy
norm; with TF32 off, the difference between the two norms is larger still (supplementary
Section~\SuppRepro{}).

In three dimensions, where the full eigendecomposition was not computed, the same quotient
was measured after the run from the norms of the errors for the networks of the
three-dimensional comparison, not for the energy-norm-supervised transformer. Its median was
\numFieldRqFree{} for the label-free transformer, \numFieldRqLab{} for the
displacement-supervised transformer and \numFieldRqMgn{} for the graph network. It was larger for the displacement-supervised
transformer than for the label-free one on
\numFieldRqAboveLab{} of the \numFieldLoads{} triples of seed, instance and load case.
Training in the Euclidean norm of the displacement thus left errors that lay higher in the
spectrum: of about the same size in two dimensions, and larger in three, where the
displacement-supervised transformer's displacement error also exceeded the label-free one's
(\cref{tab:3d}). By \eqref{eq:factor} such errors cost more energy and, by
\cref{prop:stress}, more stress error. Supplementary Section~\SuppTwo{} gives the
two-dimensional readings seed by seed, and the stress fields of one instance.
\section{Training without the solutions}\label{sec:labelfree}

\Cref{sec:norm} changed the norm and kept the labels. This section keeps the energy norm and
removes the labels: the label-free transformer is the same network trained on the energy of
the same 1,024 instances \eqref{eq:training}, without their solutions. It is compared with the usual practice, supervision on the displacement error, and with the
energy-norm-supervised transformer, from which it differs in the labels and also in the
weighting of the load cases \eqref{eq:lkgrad} and in the learning rate
(\cref{sec:methods:training}).

\subsection{Three dimensions}\label{sec:labelfree:three}

The label-free transformer was more accurate than the displacement-supervised transformer,
with or without the energy term, on all five metrics, in the seed mean and in each of the three seeds (\cref{tab:3d}). Its displacement error was
\numThreeFreeDispAdvantage{} lower, its relative energy gap \numThreeAncOverFreeGap{} times
lower than that of the variant with
the energy term, the better of the two, a factor of \numThreeAncOverFreeGapRoot{} in the
energy-norm error, and its von Mises error \numThreeFreeVm{} against \numThreeLabVm{} and
\numThreeAncVm{}. Compared instance by instance within each seed, it had the lower relative
energy gap on \numPairThreeLabGap{} of the \numThreePairs{} instance-seed pairs against the
displacement-supervised transformer, the lower von Mises error on \numPairThreeLabVm{} and
the lower displacement error on \numPairThreeLabDisp{}. The two transformers differ in the
objective and also in the learning rate, which was fixed in advance and not tuned for either
(\cref{sec:methods:training}). The pre-registered condition on the accuracy of the
label-free transformer required its displacement error to be within 10\% of the
displacement-supervised transformer's with 1,024 labels, and its relative energy gap to be
at least 40\% lower at every number of labels. It was met with a wide margin, at
\numGTwoParity{} and with a smallest advantage of \numGTwoAdvMin{}. It was one of three conditions of the comparison's gate. An amendment written after the
initial run assessed the gate's plausibility condition from 64 labels only; under that
condition's registered form, the gate would have read NO-GO. The claims here rest on the accuracy condition and on the measurements, not on the
gate's verdict (supplementary Sections~\SuppPrereg{} and~\SuppThree{}). The comparison fixed no noise guard. Read after the run with the guard of the two-dimensional
comparison, \numThreeGuardWithin{} of the \numThreeGuardTotal{} differences of seed means lie
within it, two of them, in relative energy gap and peak stress error, because of the instance
of \cref{sec:norm:three}.

Against the energy-norm-supervised transformer, which has the labels and the norm, the
label-free transformer was the less accurate. Removing the labels raised the median
energy-norm error by a factor of \numNlThreeLabelE{} and the median von Mises error by
\numNlThreeLabelVm{} (\cref{fig:normlabels}), against factors of \numNlThreeNormE{} and
\numNlThreeNormVm{} for changing the norm. The pre-registration of the energy-norm run fixed
this comparison as a reading without a criterion: the energy-norm-supervised transformer's
median relative energy gap was \numKdFreeRel{} lower than the label-free transformer's,
beyond the guard of \numKdFreeTau{}, and lower on \numKdFreePairs{} of the \numThreePairs{} instance-seed pairs. With three seeds,
the Welch 95\% interval of this reading runs from
\numKdFreeWelchLow{} to \numKdFreeWelchHigh{}. On the medians it was also the more accurate
in displacement (\numKdFreeDispRel{}), von Mises error (\numKdFreeVmRel{}) and peak stress
error (\numKdFreePeakRel{}), beyond the guard, and within the guard in critical-region
recall, readings among \numKdSecondary{} exploratory comparisons (supplementary
Section~\SuppThree{}). On the held-out instances the difference in relative energy gap was
\numKdHoldFreeRel{}, beyond a guard of \numKdHoldFreeTau{}. Which of the labels, the
weighting of the load cases and the learning rate accounts for it was not tested.

\subsection{Two dimensions}\label{sec:labelfree:two}

In two dimensions the label-free transformer was more accurate than the
displacement-supervised transformer on every metric but the displacement error, in each of the three seeds (\cref{tab:2d}). Its relative energy gap was \numCmLabOverFreeGap{}
times lower, a factor of \numCmLabOverFreeGapRoot{} in the energy-norm error, and its von
Mises error was \numCmFreeVm{} against \numCmLabVm{}. Its displacement error was
\numCmFreeDispExcessAbs{} higher, and higher in each seed. Instance by instance it had the
lower relative energy gap on \numPairCmLabGap{} of the \numCmPairs{} instance-seed pairs and
the lower von Mises error on \numPairCmLabVm{}, but the lower displacement error on only \numPairCmLabDisp{}. On \numPairCmOpposite{} of the
pairs the two errors ranked the two
networks in opposite orders (\cref{rem:rank}). The pre-registered hypothesis H1, that the
label-free transformer has the lower relative energy gap, was supported: its seed mean was
\numHOneRel{} lower, beyond the guard of \numHOneTau{}, with the same readings on the
medians and in the intervals as H2a and H2b. We have not investigated why the label-free
transformer was the more accurate in displacement in three dimensions and not in two, with
the same code.

H3, the energy-norm-supervised against the label-free transformer, was pre-registered as a
reading without a criterion, and showed no difference beyond the guard. The
energy-norm-supervised transformer's relative energy gap, \numCmKnormGap{}, was
\numHThreeRel{} lower, within the guard of \numHThreeTau{}, which its seeds made wide
(\numCmKnormGapSeeds{}, against \numCmFreeGapSeeds{}). In every seed it was the more
accurate of the two on every metric, by \numHThreeSeedRel{} in relative energy gap in seeds 0, 1 and 2. Resampling the instances, with the seeds kept, put
its relative energy gap \numHThreeResLow{} to \numHThreeResHigh{} below the label-free
transformer's (95\% interval). The guard, which treats the three seeds as the replicates, does not separate the
two networks. On the medians, removing the labels raised the energy-norm error by a factor
of \numNlCmLabelE{} and the von Mises error by \numNlCmLabelVm{}, against factors of
\numNlCmNormE{} and \numNlCmNormVm{} for changing the norm (\cref{fig:normlabels}). On the
held-out instances H3 again showed no difference beyond the guard, and the label-free
transformer's seed-mean relative energy gap was the lower, \numHoldTwoFreeGap{} against
\numHoldTwoKnormGap{} (\cref{sec:holdout}).

\subsection{Fewer labels}\label{sec:labelfree:budget}

\begin{figure}[tbp]
\centering
\figfull{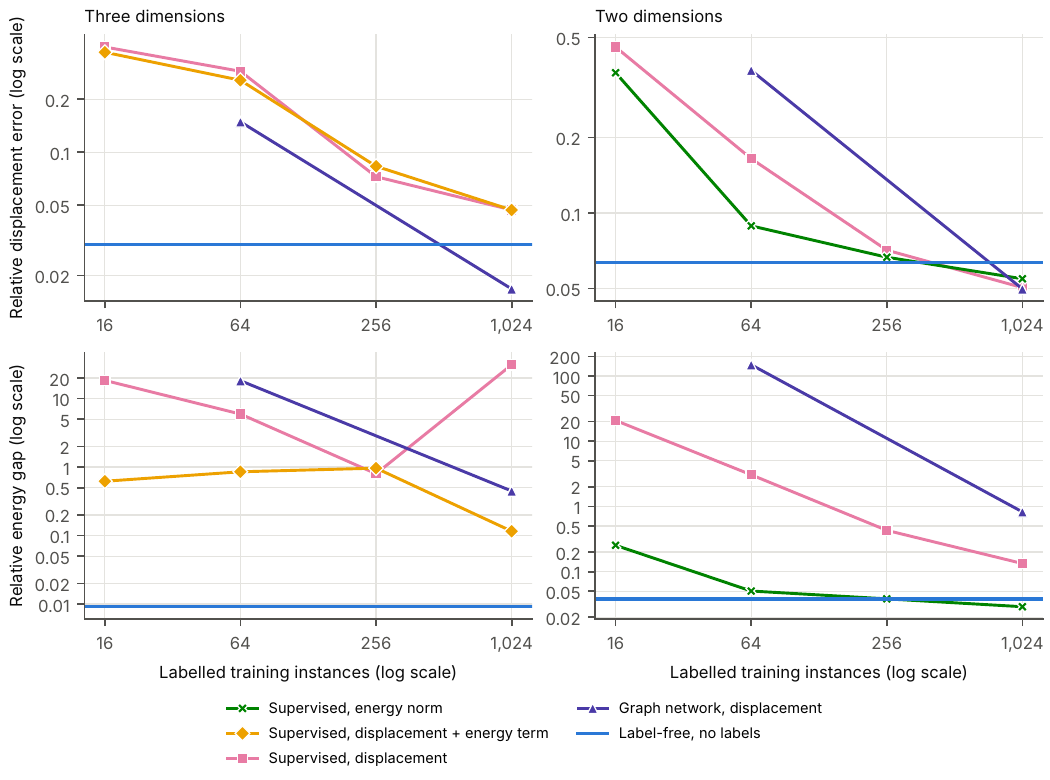}
\caption{Seed-mean relative displacement error (top) and relative energy gap (bottom) of the
supervised networks against the number of labelled training instances, in three dimensions
(left) and two (right). Horizontal line: the label-free transformer, trained on the same
1,024 instances without their solutions.}
\label{fig:labeleff}
\end{figure}

With fewer labels the supervised networks were trained for fewer steps as well
(\cref{sec:methods:training}), and their errors were in general larger (\cref{fig:labeleff};
the values are in supplementary Section~\SuppThree{}). The figure shows the
displacement-supervised transformer with the energy term in three dimensions only, and the
energy-norm-supervised transformer in two only, since in three it was trained with 1,024 labelled instances only. The graph network was
trained with 64 and 1,024 labels only. In three dimensions the label-free
transformer's relative energy gap was lower than that of every network trained on the
displacement error at every number of labels, and its displacement error lower than both displacement-supervised transformers'. The graph network reached a lower displacement error than the label-free transformer with
1,024 labels, not with 64. The displacement-supervised transformer's relative energy gap did
not fall monotonically: its seed mean rose from 256 to 1,024 labels, because of the instance of
\cref{sec:norm:three}. In two dimensions the label-free transformer's relative energy gap
was lower than those of the displacement-supervised transformer and the graph network at every number of labels. Its advantage over the displacement-supervised transformer fell from
\numCmAdvMax{} with 16 labels to \numCmAdvMin{} with 1,024. In displacement, the two
networks supervised on it reached a lower error than the label-free transformer only with 1,024 labels. Supervised in the energy norm, the transformer came within the guard of the label-free one on every
metric from 256 labels on (supplementary Section~\SuppTwo{}). With 16, 64 and 256 labelled
instances, \numCmLabWorseSixteen{}, \numCmLabWorseSixtyFour{} and \numCmLabWorseTwoFiveSix{}
of the displacement-supervised transformer's \numCmPairs{} instance-seed predictions were
worse than the zero field, against \numCmKnormWorseSixteen{}, \numCmKnormWorseSixtyFour{}
and \numCmKnormWorseTwoFiveSix{} for the energy-norm-supervised transformer and none for the
label-free one.
\section{Evaluation in the energy norm}\label{sec:checks}

The norm of the loss decides what a network learns; the norm of the metric decides how
networks are compared. This section shows that the displacement error ranked the networks
opposite to their stress errors, and tests the two checks of
\cref{sec:theory:metrics,sec:theory:detector}, which need no solution.

\subsection{Ranking by displacement and by stress}\label{sec:checks:ranking}

\begin{figure}[tbp]
\centering
\includegraphics{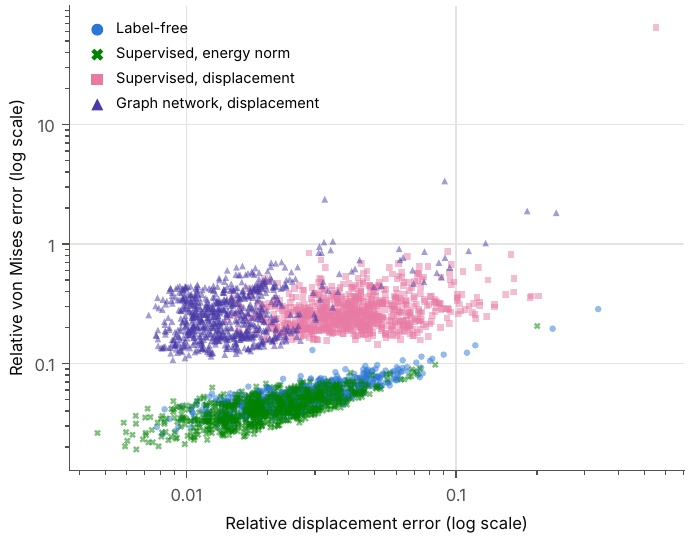}
\caption{Relative von Mises error against relative displacement error for every validation
instance and seed (\numThreePairs{} points per network): three dimensions, networks trained
on 1,024 instances.}
\label{fig:reverse}
\end{figure}

The graph network, trained on the solutions with the displacement error, had the lowest mean
displacement error of the three-dimensional networks, \numThreeMgnDisp{},
\numThreeFreeOverMgnDisp{} times lower than the label-free transformer's (\cref{tab:3d}),
and the label-free transformer was the more accurate on every other metric. The reversal held on most instances individually (\cref{fig:reverse}). The two networks
trained in the energy norm, with and without the solutions, lie at the bottom of the figure
and the two trained on the displacement error higher, whatever their displacement error. The
graph network's points lie to the left of the label-free transformer's and above them. The
point at the top right is instance No.~\numWorstIndex{} in seed \numThreeLabWorstSeed{}
(\cref{sec:norm:three}). Compared within each seed, the
graph network's displacement error was the lower on \numMgnDispLowerMin{} to \numMgnDispLowerMax{} of the instances. Its von
Mises error, however, was the higher on
\numMgnVmHigher{} of the \numThreePairs{} instance-seed pairs, with a mean
\numThreeMgnOverFreeVm{} times higher, and its relative energy gap was the higher on at
least \numMgnGapHigherMin{} of the instances in every seed. On \numMgnLowerDispHigherVm{}
pairs both held at once. The peak stress error, which compares a single value per load case,
followed the pattern less closely: the graph network's was the lower on
\numMgnPeakLowerMin{} to \numMgnPeakLowerMax{} of the instances, although its mean was
\numThreeMgnOverFreePeak{} times higher. The relative energy gap thus ranked the two
networks as the von Mises error did, not as the displacement error did. In two dimensions
the graph network again had the lowest displacement error of the networks, \numCmMgnDisp{},
within the guard of the displacement-supervised transformer's. Its von Mises error was
\numCmMgnOverFreeVm{} times the label-free transformer's, and its relative energy gap the
largest of the trained networks, \numCmMgnOverLabGap{} times the displacement-supervised
transformer's (\cref{tab:2d}). By \cref{rem:rank} such reversals are possible whenever the
displacement errors differ by a factor of less than $\kappa^{1/2}$, which on the instances
measured was at least about \numSqrtKappaMin{} (\cref{sec:methods:data}). These comparisons use the solutions. For the predictions of one load case, differences of energy rank them
exactly as the relative energy gap does, without a solution (\cref{cor:rank}).

\begin{figure}[tbp]
\centering
\figfull{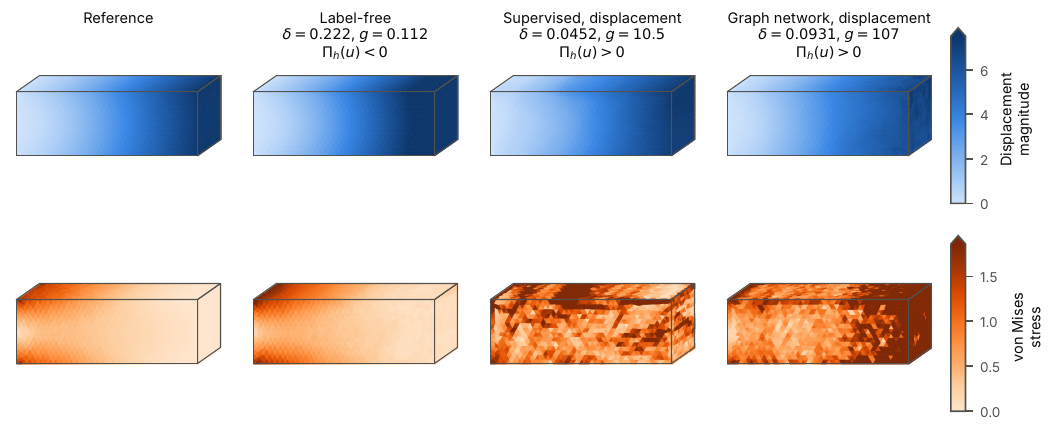}
\caption{Validation instance No.~\numWorstIndex{}, seed 0, under the \numFigWorstLoad{}:
displacement magnitude (top) and von Mises stress (bottom) of the reference solution and of
three networks trained on 1,024 instances, on the three visible faces of the box; one colour
scale per row, set by the reference (values above its maximum saturate). Above each
network: its relative displacement error $\delta$, its relative energy gap $g$ and the sign
of its potential energy $\Pi_h(u)$ on this load case (positive: worse than the zero field,
\cref{cor:zero}).}
\label{fig:field-worst}
\end{figure}

\Cref{fig:field-worst} shows the pattern on one instance, No.~\numWorstIndex{} in seed 0,
the one with the largest relative energy gap of the displacement-supervised transformer
(selection rules in supplementary Section~\SuppRepro{}).
Averaged over its four load cases, the label-free transformer had the largest displacement
error of the three networks, \numWorstFreeDisp{}, and the most accurate stresses, with a von
Mises error of \numWorstFreeVm{} and a relative energy gap of \numWorstFreeGap{}. The
displacement-supervised transformer and the graph network had displacement errors of
\numWorstLabDisp{} and \numWorstMgnDisp{}, von Mises errors of \numWorstLabVm{} and
\numWorstMgnVm{}, and relative energy gaps of \numWorstLabGap{} and \numWorstMgnGap{}: both
were more accurate in displacement and worse than the zero field in energy. On the load case
shown, the three displacement fields have the shape of the reference, the label-free one
with too large an amplitude: its relative displacement error of \numFigWorstFreeDisp{} falls
to \numFigWorstFreeDispC{} when it is rescaled by $\cstar=\numFigWorstFreeC{}$
(\cref{prop:amp}). The stress fields of the two networks trained on the displacement error
are rough, and more than twice the reference's on \numFigWorstTwiceLab{} and
\numFigWorstTwiceMgn{} of the solid's volume, against \numFigWorstTwiceFree{} for the label-free transformer. A typical instance, on which their predictions were better than the
zero field, shows the same roughness (supplementary Section~\SuppThree{}). Small
displacement errors rich in stiff modes are what \cref{sec:norm:spectrum} measured, and by
\cref{prop:stress} they are large stress errors.

\subsection{Predictions worse than the zero field}\label{sec:checks:zero}

\begin{figure}[tbp]
\centering
\figfull{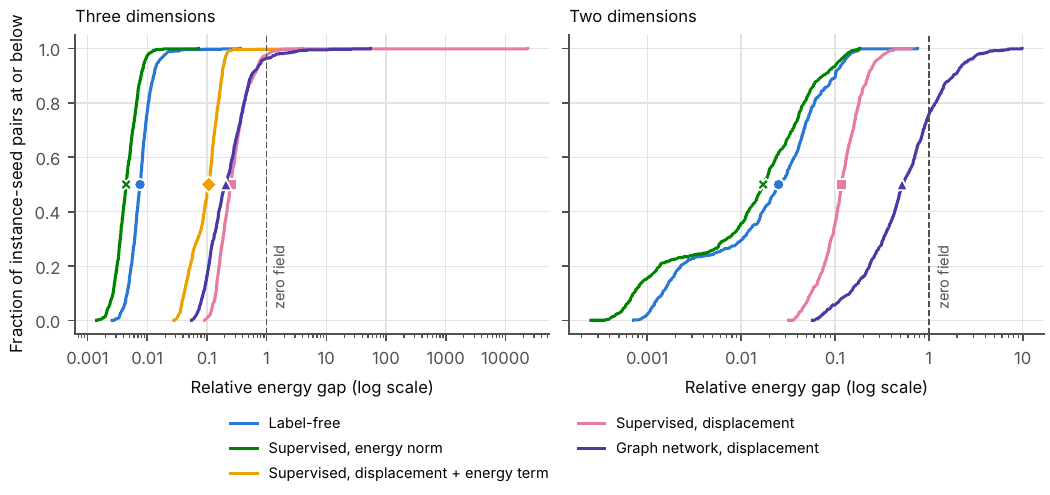}
\caption{Distribution of the relative energy gap over the \numThreePairs{} validation
instance-seed pairs, each the mean over the instance's four load cases: networks trained on
1,024 instances, in three dimensions (left) and two (right). Markers: medians; right of the
dashed line: worse than the zero field.}
\label{fig:energygap}
\end{figure}

In three dimensions none of the \numThreePairs{} instance-seed predictions of the label-free
or of the energy-norm-supervised transformer had a mean relative energy gap above one
(\cref{fig:energygap}), against \numThreeLabWorse{} of the displacement-supervised
transformer's, \numThreeAncWorse{} of the variant with the energy term and \numThreeMgnWorse{} of the graph network's. The label-free transformer's median over the
pairs, \numThreeFreeGapMedian{}, was \numThreeMedianRatioMin{} to \numThreeMedianRatioMax{}
times lower than the medians of the networks trained on the displacement error. In two
dimensions \numCmMgnWorse{} of the graph network's predictions had a mean relative energy
gap above one, and none of any transformer's. These counts use the solutions. Load case by
load case, with the test of \cref{cor:zero}, which does not, \numFieldLoadWorseLab{} of the
displacement-supervised transformer's \numFieldLoads{} three-dimensional predictions and
\numFieldLoadWorseMgn{} of the graph network's had a positive energy, and so were worse than
the zero field, against \numThreeFreeLoadWorse{} of the label-free transformer's. In two dimensions the counts
were \numSpecLoadWorseMgn{} of the graph network's \numSpecLoads{}, against at most one of
each transformer's. These counts were made after the runs. On the held-out instances
the same test was a pre-registered claim. Of the \numHoldTThreeLoads{} load cases of each of
the two networks trained on the displacement error in three dimensions,
\numHoldTThreeLab{} and \numHoldTThreeMgn{} were worse than the zero field; these were
exactly the load cases with a positive energy (\cref{tab:holdout}). After the rescaling by $\cstar$, none of the three-dimensional
predictions flagged was worse than the zero field, and the rescaling never increased a
relative energy gap. The failures of the displacement-supervised transformer and of the
graph network recurred: more than half of the instances on which either had a mean relative
energy gap above one had it in two or three of its seeds (supplementary
Sections~\SuppThree{} and~\SuppTwo{}).

The check needs only the assembled $K$ and $F$. Where a surrogate is used without them,
the check and the rescaling by $\cstar$ cost the assembly of $K$ and $F$, which here took longer than the
surrogate's inference (\cref{tab:cost}). The check applies to any network, however it was
trained, and the rescaling, also without a solution, makes every prediction that the check
flags no worse than the zero field (\cref{prop:amp}). The check catches gross failures only: the displacement-supervised transformer's median prediction in
three dimensions, with a relative energy gap of \numThreeLabGapMedian{}, passed it.
\section{Held-out instances}\label{sec:holdout}

\begin{table}[tbp]
\centering
\footnotesize
\setlength{\tabcolsep}{4pt}
\caption{The claims of the comparisons that the pre-registered holdout evaluation read again on the held-out instances (\cref{sec:holdout}), with the same stored networks and as its adjudicator recorded them, beside their values on the validation instances. T1 and D2: comparisons that hold, five metrics in each seed and in the seed mean; T2: the range of the shares over the seeds, then the share of the instance-seed pairs; H1 to H3: the relative difference rel and its noise guard (\ref{app:claims}).}
\label{tab:holdout}
\resizebox{\ifdim\width>\textwidth\textwidth\else\width\fi}{!}{%
\begin{tabular}{@{}>{\raggedright\arraybackslash}p{0.07\linewidth}>{\raggedright\arraybackslash}p{0.455\linewidth}>{\raggedright\arraybackslash}p{0.14\linewidth}>{\raggedright\arraybackslash}p{0.14\linewidth}>{\raggedright\arraybackslash}p{0.155\linewidth}@{}}
\toprule
Claim & Statement & Validation & Holdout & Outcome \\
\midrule
T1 (3D) & Label-free transformer better than the displacement-supervised transformer on all five metrics, each seed and the seed mean & 20 of 20 & 20 of 20 & replicated \\
T2 (3D) & Graph network: lower displacement error than the label-free transformer on most instances, each seed; higher von Mises error on at least 95\% of the pairs & 84.4\%--92.6\%; 99.9\% & 86.3\%--95.8\%; 100\% & replicated \\
T3 (3D) & Load cases worse than the zero field, displacement-supervised transformer and graph network, each with a positive energy & 126 and 106 of 3,072 & 340 and 237 of 8,232 & replicated \\
\midrule
H1 (2D) & Label-free below the displacement-supervised transformer in relative energy gap, beyond the guard & $-$71.3\% (guard 12.4\%) & $-$71.3\% (guard 11.5\%) & replicated \\
H2a (2D) & Energy-norm-supervised below the displacement-supervised transformer in relative energy gap, beyond the guard & $-$78.2\% (guard 16.4\%) & $-$42.9\% (guard 47.4\%) & \hspace{0pt}not replicated \\
H2b (2D) & The same in von Mises error & $-$54.9\% (guard 12.3\%) & $-$52.8\% (guard 12.9\%) & replicated \\
H2c (2D) & The same, by more than half & $-$54.9\% & $-$52.8\% & replicated \\
D2 (2D) & Label-free transformer: higher displacement error than the displacement-supervised one, lower errors on the other four metrics, each seed and the seed mean & 20 of 20 & 20 of 20 & replicated \\
H3 (2D) & A reading: energy-norm-supervised against label-free transformer, relative energy gap & $-$24.2\% (guard 38.1\%) & +99.1\% (guard 160.2\%) & no difference shown \\
\midrule
V & Per-instance values alike on the holdout and validation instances, 35 quantities & -- & none rejected & \hspace{0pt}representative \\
\bottomrule
\end{tabular}}
\end{table}

The comparisons of \cref{sec:norm,sec:labelfree,sec:checks} were read on the validation
instances, on which the runs were monitored and, in two dimensions, on which the results
that motivated the comparison had been seen. A pre-registered evaluation read eight of their
claims again, with the stored networks unchanged, on instances of the same corpora on which
no recorded run or tool had trained or evaluated a network or computed a label. There were \numHoldThreeN{} such instances in three dimensions and
\numHoldTwoN{} in two
(\cref{sec:methods:data}), labelled only after the rules had been fixed. The
displacement-supervised transformer with the energy term, whose networks were not stored,
was not evaluated. \Cref{tab:holdout} gives the claims with their values on both sets; seven
of the eight replicated. The verdicts are not independent: the claims of each dimension read
the same instances and several of them the same networks, and agreement among them adds
little evidence.

In three dimensions the label-free transformer was again more accurate than the
displacement-supervised transformer on all five metrics, in each seed and in the seed mean. The graph network again had the lower displacement error on
most instances and the higher von Mises error, now on every instance-seed pair
(\numMgnVmHigher{} of \numThreePairs{} on the validation instances). The networks trained on the displacement error again had load
cases worse than the zero field, each with a positive energy. In two dimensions the
label-free transformer's lower relative energy gap, and its larger displacement error with
better values of the other four metrics, replicated, as did the energy-norm-supervised
transformer's lower von Mises error, beyond the guard and by more than half.

The energy-norm-supervised transformer's lower mean relative energy gap in two dimensions
did not replicate. Its seeds' relative energy gaps on the held-out instances were
\numHoldHTwoANewSeeds{}, against \numHoldHTwoABaseSeeds{} for the displacement-supervised
transformer: \numHoldHTwoARel{} lower, within the guard of \numHoldHTwoATau{}. The readings
fixed beside it locate the failure in a few instances. Its mean relative energy gap was
\numHoldHTwoARatio{} times as large on the held-out as on the validation instances (95\%
interval, the held-out instances resampled: \numHoldHTwoARatioLow{} to
\numHoldHTwoARatioHigh{}), whereas its median was much the same on both,
\numHoldHTwoAMedHold{} and \numHoldHTwoAMedVal{}. Of its \numHoldTwoPairs{} instance-seed
predictions on the held-out instances, \numHoldTwoWorseInstKnorm{} had a mean relative
energy gap above one, against \numHoldTwoWorseInstLab{} of the displacement-supervised
transformer's and \numHoldTwoWorseInstFree{} of the label-free transformer's; on the validation instances none of the three had any. Read
after the evaluation, the largest
relative energy gaps of its three seeds were \numHoldHTwoATopSeeds{}. Its predictions above
one lay on as many instances, each in one seed only, and its other two seeds predicted those
instances with relative energy gaps of at most \numTailTwoKnormOther{} (supplementary
Section~\SuppTwo{}). On the seeds' medians
the reduction was \numHoldHTwoAMedRel{}, beyond a guard of \numHoldHTwoAMedTau{}: the claim
holds for the typical instance, not for the mean, which a few held-out instances dominate. H3
again showed no difference beyond the guard, and the label-free transformer's seed-mean
relative energy gap was now the lower, \numHoldTwoFreeGap{} against \numHoldTwoKnormGap{}.

Beside the claims, the per-instance values of every network and metric were compared between the two sets. Of the \numHoldVQuantities{} quantities none was rejected by two-sided
Mann--Whitney tests of the held-out against the validation values, taken together by Holm's
procedure at a family-wise level of \numHoldVAlpha{}, and the validation instances were read
as representative. That verdict does not show that the means are unchanged: a mean that a few
instances dominate can move without a rejection, as the energy-norm-supervised transformer's
relative energy gap did. Supplementary Section~\SuppPrereg{} gives the means and medians on
both sets and the evaluation's two departures from its plan.
\section{A classical baseline and the cost}\label{sec:cost}

\subsection{The Ritz projection}\label{sec:cost:ritz}

The Ritz projection of \cref{sec:methods:ritz} needs the instance's stiffness matrix but no
labels or training (its rows in \cref{tab:3d,tab:2d}). On the 256 three-dimensional
validation instances its displacement error was smaller than the label-free transformer's on
\numRitzValFreeDispSixMin{} to \numRitzValFreeDispSixMax{} of the instances at $p=6$,
depending on the seed, and on \numRitzValFreeDispNineMin{} to \numRitzValFreeDispNineMax{} at $p=9$. At $p=9$ it was
smaller than each network's, in each seed, on at least
\numRitzValNineDispAll{} of the instances. Its median relative energy gap fell from
\numRitzValGapSix{} at $p=6$ to \numRitzValGapNine{} at $p=9$, against \numKdFreeMedMin{} to
\numKdFreeMedMax{} for the label-free transformer's three seeds. At $p=9$ its seed means were better than the label-free transformer's on every metric but the peak stress error. Its
median peak stress error, \numRitzValPeakSix{} at $p=6$ and \numRitzValPeakNine{} at $p=9$,
was \numRitzValPeakRatioMin{} to \numRitzValPeakRatioMax{} times the label-free
transformer's, and its peak stress error was the larger on at least \numRitzValFreePeakMin{}
of the instances at every degree. The energy-norm-supervised transformer had the lower
relative energy gap on at least \numRitzValKnormGapMin{} of the instances and the lower peak stress error on all of them, at every degree. It also had lower seed means
than the projection at
$p=9$ in relative energy gap, von Mises error and peak stress error. On the held-out
instances the comparisons gave much the same result: the energy-norm-supervised transformer had the
lower relative energy gap on at least \numRitzHoldKnormGapMin{} of them and the lower peak
stress error on all but at most \numRitzHoldKnormPeakMiss{}, at every degree. In two
dimensions, with \numRitzTwoKSix{} and \numRitzTwoKEight{} unknowns, the projection's
displacement error was smaller than the label-free transformer's on \numRitzTwoFreeDispMin{}
to \numRitzTwoFreeDispMax{} of the instances, depending on the seed and the degree. Its relative energy gap was smaller on \numRitzTwoFreeGapSixMin{} to \numRitzTwoFreeGapSixMax{} at
$p=6$ and on \numRitzTwoFreeGapEightMin{} to \numRitzTwoFreeGapEightMax{} at $p=8$, and its
peak stress error larger on at least \numRitzTwoFreePeakMin{} of them. Supplementary
Section~\SuppRitz{} gives the shares for every network, set and degree. A smooth polynomial
space of a few hundred unknowns thus approximates the displacement field of these solids
better than each network, and their largest stresses worse than the transformers trained in
the energy norm.

\subsection{Inference and solving}\label{sec:cost:inference}

\begin{table}[tbp]
\centering
\footnotesize
\setlength{\tabcolsep}{4pt}
\caption{Wall time per instance of the label-free transformer's inference, on one GPU, and of solves of the same systems on the same machine's CPU, all load cases together and without the assembly, with the surrogate's accuracy. CG: unpreconditioned conjugate gradients; its iterations (from the zero field / to the prediction's accuracy) and its displacement error are medians over the load cases, the other times and errors medians over the instances. $c^*$: energy-optimal amplitude \eqref{eq:cstar}. Protocol in supplementary Section~\SuppTiming{}.}
\label{tab:cost}
\resizebox{\ifdim\width>\textwidth\textwidth\else\width\fi}{!}{%
\begin{tabular}{lccc}
\toprule
 & 2D & 3D & 3D, fine mesh \\
\midrule
Instances (load cases) & 32 (128) & 32 (128) & 8 (32) \\
Nodes, median & 562 & 4,960 & 38,156 \\
\midrule
Surrogate, first call & 3.2 ms & 49 ms & 2.15 s \\
Surrogate, repeated call & 2.4 ms & 49 ms & 2.16 s \\
\midrule
Sparse direct solve & 3.8 ms & 1.22 s & 133 s \\
CG from the zero field & 28 ms & 1.72 s & 26.1 s \\
CG from the prediction & 29 ms & 1.71 s & 27.4 s \\
CG to the prediction's accuracy & 8.8 ms & 292 ms & 4.50 s \\
the same, after rescaling by $c^*$ & 8.7 ms & 296 ms & 5.01 s \\
\midrule
Assembly of $K$, the repository's assembler & 9.5 ms & 900 ms & 8.07 s \\
Ritz projection after assembly & 1.4--2.1 ms & 34--144 ms & 0.54--1.39 s \\
\midrule
CG iterations per load case & 290 / 85 & 746 / 134 & 1,456 / 241 \\
\midrule
Surrogate's relative energy gap & 0.029 & 0.0085 & 0.12 \\
Surrogate's displacement error & 0.050 & 0.031 & 0.27 \\
Relative energy gap after rescaling by $c^*$ & 0.028 & 0.0077 & 0.065 \\
Displacement error after rescaling by $c^*$ & 0.040 & 0.014 & 0.076 \\
CG's displacement error at that accuracy & 0.049 & 0.017 & 0.16 \\
\bottomrule
\end{tabular}}

\smallskip
\begin{minipage}{\linewidth}\footnotesize\raggedright
Fine mesh, direct solve: the median over the seven of the eight instances that it solved before its budget of 900 s ran out.\par
CG to the prediction's accuracy uses the reference solution, so it is a lower bound; on the fine mesh it needs no iteration on the three load cases whose relative energy gap exceeds 1.
\end{minipage}
\end{table}

\Cref{tab:cost} compares the time of the label-free transformer's inference with that of
solving the same systems on the same instances, with the sparse direct solver that produced the labels and with conjugate gradients. The network ran on the GPU and the solvers on the CPU of the
same machine. The assembly of the stiffness matrix, which the solvers and the Ritz projection
need and the surrogate does not, is listed apart (supplementary Section~\SuppTiming{} gives
the protocol). In three dimensions the direct solve took
a median \numCostThreeValDirectRatio{} times as long as the surrogate's first call on the validation instances, and
\numCostThreeFineDirectRatio{} times as long on the \numCostFineDirectSolved{} fine-mesh
instances that it solved before its time budget ran out (here and below, medians of the
per-instance ratios, not ratios of the medians in \cref{tab:cost}). On those fine-mesh
instances the surrogate's median relative energy gap was \numCostThreeFineGap{}, against
\numCostThreeValGap{} on the validation instances. In two dimensions the direct solve took
\numCostTwoDirectRatio{} times as long, and its solution is exact. Conjugate gradients from
the zero field, stopped at the first iterate as accurate as the prediction in the energy
norm, took a median \numCostThreeValMatchRatio{} and \numCostThreeFineMatchRatio{} times the
surrogate's first call on the three-dimensional validation and fine-mesh instances. This stop uses the reference
solution, so a practical run would take at least as long. The Ritz
projection took \numCostRitzValMin{} to \numCostRitzValMax{} per validation instance from
$p=6$ to $p=9$, against \numCostSurrVal{} for the surrogate's first call, after an assembly
that took \numCostAsmOverRitzMin{} to \numCostAsmOverRitzMax{} times as long as the
projection itself with the repository's assembler. These times do not show a speed advantage of the surrogate. The network ran on a GPU and the
solvers on a CPU, with \numCostThreads{} threads (supplementary Section~\SuppTiming{}).
Neither the general sparse LU factorisation of SuperLU nor unpreconditioned conjugate
gradients is the solver of choice for these symmetric positive definite systems. The ratio of
the matched solve to the surrogate also fell from \numCostThreeValMatchRatio{} to
\numCostThreeFineMatchRatio{} on the finer mesh.

\subsection{Labels and training}\label{sec:cost:labels}

In three dimensions the comparison consumed \numThreeSolves{} solves: \numThreeTrainSolves{}
for the 1,024 labelled training instances, \numThreeFewSolves{} for 64 fine-mesh instances
reserved for few-shot training and \numThreeEvalSolves{} for the in-band and fine-mesh validation sets. In two dimensions the comparison used
\numCmSolves{}, \numCmSolvesTrain{} of them for
training. The validation solves are needed to evaluate any method; the training solves are
the price of supervision, and the label-free transformer needed none of them. Here they cost
little. At the median times of \cref{tab:cost}, the \numThreeTrainSolves{} three-dimensional
training solves take about \numCostLabelMin{} min of the direct solver, against
\numTimeThreeSeed{} h of training per seed on one NVIDIA GeForce RTX 5090, for 204,800
steps, which the displacement-supervised transformer needs as well. For these problems,
label-free training saves those minutes, not the cost of training. Full self-attention over
all nodes dominates that cost and grows with the square of the number of nodes: one training
step took \numEtwoTransformerFineStep{} s on a fine-mesh instance of \numEtwoNodesFine{}
nodes. A variant that attends over pooled tokens trained \numTimeBottleneckSpeedup{} times
faster but was less accurate, and was dropped under its pre-registered criterion
(supplementary Section~\SuppSide{}). Set against the direct solve, the training of one seed
is recovered after about \numCostBreakEvenDirect{} predicted instances, and after about
\numCostBreakEvenMatch{} against conjugate gradients stopped at the surrogate's accuracy, an
upper bound since that stop uses the reference solution. Label-free training recovers its
cost where a solve is expensive enough
(\cref{sec:discussion:engineering,sec:discussion:outlook}). As warm starts the
predictions did not shorten conjugate gradients, here or in the pre-registered test of the
July run (K5 in \cref{tab:claims}). Supplementary Section~\SuppWarm{} gives these measurements, the iteration bound that explains them and a comparison with reported
speed-ups.
\section{Discussion}\label{sec:discussion}

\subsection{The norm, not the labels}
Both comparisons changed the norm of the supervised loss and nothing else. Both lowered the
stress error, on the validation and on the held-out instances: in three dimensions the
median von Mises error by a factor of \numNlThreeNormVm{}, in two the mean by more than half
(\cref{sec:norm}). In two dimensions, measured after the run, the errors of the two
transformers were of about the same size and lay at different places in the spectrum.
Removing the solutions with the energy norm kept raised the median energy-norm and von Mises
errors by factors of at most \numNlThreeLabelWithin{} (\cref{sec:labelfree}), in a comparison
that also changes the weighting of the load cases and the learning rate. The label-free
transformer's advantage over displacement supervision thus lay in the norm of its loss rather
than in the absence of labels. By \cref{cor:training,prop:stress} the energy objective is, up
to a constant, the compliance-weighted mean-square error of the stress field, which is what
the displacement error under-weights. With labels at hand, training in the energy norm with
them was the more accurate in three dimensions and showed no difference beyond the guard in two. The label-free objective is thus not necessarily
the most accurate way to train in the
energy norm, but it is the way to train in it without solutions. The energy term of the
displacement-supervised transformer, which adds the energy to the displacement loss, is a step towards the energy norm. It removed most of the large failures, and the label-free
transformer, which has only that term, was more accurate still (\cref{tab:3d}).

\subsection{Engineering use}\label{sec:discussion:engineering}
For surrogates used in stress analysis, such as those of the families of
\cref{sec:methods:data}, the results suggest three rules. First, train in the energy norm,
which by \cref{cor:training,prop:stress} is, up to a constant, the compliance-weighted error
of the stresses, so a surrogate trained in it is in effect supervised on its stresses.
Given the assembled system, the energy norm of an error costs one sparse matrix-vector
product per load case. A supervised loss can therefore be measured in it with the labels
already computed, on which it depends only through the product $F^{\top}\Ustar$ of each load case (\cref{sec:methods:training}). Where no solutions have been computed, minimising the
energy trains in the same norm without them. Second, compare surrogates by their energy-norm and stress errors, not by their displacement
errors alone. Surrogates compared by displacement error alone can be ranked opposite to their
stresses (\cref{sec:checks:ranking}), and an evaluation can report the energy norm beside the
displacement error. Third, screen
predictions with the energy before they are used. Its sign flags gross failures, those worse
than the zero field, without a solution (\cref{sec:checks:zero}). Each of the \numHoldTwoWorseInstKnorm{}
predictions of the energy-norm-supervised transformer with a mean
relative energy gap above one on the two-dimensional held-out instances (\cref{sec:holdout})
had a load case that the screen flags. Differences of energy rank the predictions of a load case
exactly (\cref{cor:rank}), so, of several predictions from different networks or seeds
or from a network and a Ritz projection, the one of lowest energy is the most accurate in the
energy norm. A prediction that fails should be sent back to the solver. Rescaling it by
$\cstar$ (\cref{prop:amp}) makes it no worse than the zero field without a solve, but not
necessarily accurate.

The gain is in the accuracy of the stresses, not in speed. Training in the energy norm leaves
the cost of a prediction unchanged. Conjugate gradients given the stiffness matrix, stopped by
a test that uses the reference solution, reached the label-free transformer's accuracy within a few
times its inference time (\cref{sec:cost:inference}). Training without labels saves only their
solves, which matters where those are expensive (\cref{sec:cost:labels}). Surrogates, and with them these
rules, are worth their training only where one family of components is analysed many times.
In three dimensions the training of one network was recovered against the direct solve after
about \numCostBreakEvenDirect{} predicted instances (\cref{sec:cost:labels}), with the
caveats of \cref{sec:cost:inference}, under which a faster solver would raise that number. In two dimensions the direct solve took \numCostTwoDirectRatio{} times as long as the inference
and is exact, so a surrogate of plates of this size saves little or no time. The screen needs the stiffness matrix and the load vectors of the instance. Their assembly
took a median of \numCostAsmVal{} against \numCostSurrVal{} for the inference in three
dimensions (\cref{tab:cost}). It is also most of the cost of a Ritz projection, which took a further
\numCostRitzValMin{} to \numCostRitzValMax{} and, at the top of its ladder, was more accurate
than the label-free transformer, in the seed means, on every metric but peak stress
(\cref{sec:cost:ritz}). Passing the screen does not make a prediction accurate
(\cref{sec:checks:zero}), and the accuracy of a surrogate on a family has to be established on
solved instances of that family, as the validation instances establish it here.

In the assessment of parts whose internal voids differ from one part to the next, each part is a new instance of a family. The solids here, with at most three spherical cavities under
four load cases, are a stylised version of that setting. Real voids are more numerous, smaller and irregular. Resolving them needs meshes finer than any on which these networks
were accurate (\cref{sec:discussion:transfer}), and other geometries, numbers of voids and
loads were not tested. Label-free training would be most useful where each solve is expensive, as
in the nonlinear and time-dependent problems of \cref{sec:discussion:outlook}.

As in an energy-based model
\cite{LeCun2007EBM}, which uses one scalar energy to find the most compatible output, to rank
candidates and to detect incompatible ones, the potential energy here is used to train, to rank and to check. It is assembled rather than learned, and exact. The identity does not
depend on the approximation: the energy evaluates a Ritz solution as it does a network's prediction. The rescaling by $\cstar$, the projection onto the span of a prediction
(\cref{prop:amp}), is the smallest enrichment of a finite-element space by it; richer
enrichments have been analysed \cite{Barucq2026Enrich}. Where the stiffness matrix is not at
hand, as for solutions exported from another code, a loss on the strains or the stresses is
the natural substitute; we did not test one.

\subsection{Earlier comparisons}\label{sec:discussion:earlier}
The account is consistent with part of the earlier reports. In soft-tissue beams, adding the
deformation gradient to the supervised loss, a step towards a norm of the strains, narrowed
the gap between supervised and label-free training \cite{Dalton2023}. For a scalar multiscale
problem an energy-norm loss gave lower errors of the solution than $L^2$,
relative $L^2$ or $L^1$ losses on the same labels \cite{Haltmayer2026LOD}. One comparison,
reported in a preprint, points the other way. In the pressure vessels of Sun
et~al.~\cite{Sun2026FEVessel}, the same point-cloud transformer trained on the relative
error of the displacement labels instead of the energy, with every other setting unchanged,
was more accurate in displacement and in the $H^1$ seminorm (their Table 2), against the
three-dimensional result of \cref{sec:labelfree:three}. The two studies differ in ways that
may matter. That comparison is a single run, and its reported errors are the lowest test
errors over the last third of training. Its reference solutions use reduced-integration
hexahedra with hourglass control, and its energy loss takes derivatives from the shape
functions of those elements, with a quadrature that is not stated. \Cref{lem:exact} makes
energy training exact regression on the reference solution only when the energy minimised is
the assembled energy of the discretisation that defines the reference. If the two differ,
the minimiser of the energy loss is not the reference solution, and energy training
regresses towards a different target. Whether this explains the reversal is not known. The
comparison shows that the advantage of label-free training over displacement supervision
reported here is not universal.

\subsection{Transfer}\label{sec:discussion:transfer}
No network transferred to the finer mesh. The readings of supplementary
Section~\SuppTransfer{} point to the output scaling. The output is multiplied by the largest nodal load, which depends
on the mesh while the solution does not. The networks of the July run, whose output was not scaled in this way, lost little accuracy on meshes
\numCoarsenLow{} and \numCoarsenHigh{} times finer than those they were trained on, though
that evidence is weak. Until an output scale independent of the mesh is shown to remove the
failure, the label-free transformer of this paper should be regarded as valid within the range of mesh sizes it was trained on. The discretisation invariance reported for
point-cloud operators trained on the energy \cite{Wang2026PretrainFEM,Sun2026FEVessel} was
tested by varying the density of the input point cloud rather than by refining the mesh.
Label-free fine-tuning on the finer mesh, which needs no solves, was not tested here.

\subsection{Limitations}
The study is restricted to linear elastostatics, where \cref{lem:exact} is exact; one family
of geometries and one set of four load cases per dimension; one model size; and three seeds
per network, for which the noise guard is coarse. \Cref{prop:stress} assumes a homogeneous
isotropic material and an exactly integrated stiffness, as with the linear elements used here. The reported von Mises error weights elements equally, so the proposition
explains it rather than bounds it, although the area- and volume-weighted errors, measured
after the runs, differed little. None of the metrics isolates the stresses at the holes or cavities,
and where the peak of the discrete solution lies at the clamped edge, the peak stress
error does not measure the concentration at a hole (\cref{sec:methods:metrics}). Only the transformer was trained in the energy norm. The graph network was trained on the displacement error alone, and whether its rough stress
fields come from the norm or from message passing over the mesh edges was not tested. The
learning rates were fixed before the runs and not tuned, and the energy-norm-supervised
transformer used the displacement-supervised settings unchanged. The departures from the
pre-registrations are stated with their results: the statistic of the three-dimensional
energy-norm run (\cref{sec:norm:three}), the registration of the two-dimensional comparison
(\cref{sec:norm:two}), the amended three-dimensional gate (\cref{sec:labelfree:three}) and the claim that did not replicate (\cref{sec:holdout}). The spectral readings and the Ritz
baseline were added after the runs. A pre-registered comparison kept the joint-embedding term, a self-supervised term in which
the network predicts the latent vectors of masked regions of the mesh. The scope of this paper, the energy
objective without that term, was chosen after that comparison (supplementary
Section~\SuppPrereg{}). The cost of full self-attention grows with the square of
the number of nodes, which limits the size of the meshes that the transformer can take.

\subsection{Outlook}\label{sec:discussion:outlook}
The construction extends wherever each step of a computation is a convex minimisation. In
linear dynamics, every implicit step of a Newmark-type integrator solves a symmetric
positive definite system whose effective stiffness combines the stiffness, mass and damping
matrices, and \cref{lem:exact} holds step by step \cite{CaoSong2026}. Incremental
variational formulations of dissipative materials write each step of a nonlinear history as
the minimisation of an incremental potential \cite{OrtizStainier1999}, and variational
integrators derive each step of a time integration from a discrete variational principle
\cite{MarsdenWest2001}. Where such a step is a minimisation, its potential remains a
label-free training objective, although the energy gap is no longer a squared norm, and the
counterparts of \cref{prop:amp,cor:zero} remain to be worked out. At finite strains the incremental potential is in general not convex \cite{OrtizStainier1999}, so its
minimiser need not be unique. These are the problems in which a solve may be expensive
enough for label-free training to recover its cost, and in which the choice of norm may again
decide what a surrogate learns.
\section{Conclusion}\label{sec:conclusion}

For linear elastostatics, the assembled discrete potential energy exceeds its minimum by half
the squared energy-norm error of a prediction, which is the strain energy of its stress error.
Minimising it is supervised regression in the energy norm without solutions; the usual
displacement loss regresses in the Euclidean norm, which under-weights the stiff modes that
carry the strains. In pre-registered comparisons that changed only the norm of the
supervised loss, with the same solutions, the median von Mises error of a three-dimensional
transformer fell \numNlThreeNormVmWord{}fold, and in two dimensions the mean fell by more than half, also on held-out instances. In two dimensions, measured after the run, the errors of the
displacement-supervised and the energy-norm-supervised transformers were of about the same
size, but the latter's lay lower in the spectrum of the stiffness matrix. Removing the solutions, with the energy norm
kept, cost far less accuracy. In three dimensions the
transformer
trained on the energy alone came within a factor of \numNlThreeLabelWithin{} of the
energy-norm-supervised one in median energy-norm and von Mises error, and was more accurate
than the displacement-supervised one on every metric, in every seed. For these surrogates,
the norm of the loss, not the use of solutions, limited the stress accuracy. The same norm
should decide how surrogates are compared. A graph network with the lowest displacement error
of the networks had among the largest stress errors. Without a solution, the energy itself
ranks the predictions of a load case and, by its sign, flags those worse than the zero field.
Given the stiffness matrix, a polynomial Ritz projection of a few hundred unknowns was more
accurate than the label-free transformer, in the seed means, on every metric but peak
stress. No network transferred to a substantially finer mesh. These limits, the
claim that did not replicate on held-out instances and the criteria that were not met are
reported with the results. For engineering use, these results suggest training surrogates
for stress analysis in the energy norm, comparing them by their energy-norm and stress
errors rather than by their displacement errors alone, and screening each prediction for
gross failures by the sign of its energy before it is used.

\appendix
\section{Pre-registered claims and their outcomes}\label{app:claims}
\setcounter{table}{0}

\Cref{tab:claims} lists the pre-registered claims of the three comparisons of this paper
(\cref{sec:methods:prereg}), and the pre-registered test of a warm start, with the values measured and the outcomes as the code recorded them at the time.
\Cref{tab:holdout} gives
those of the holdout evaluation. The hypotheses of the comparisons are evaluated in
\cref{sec:norm,sec:labelfree}, and the gate of the three-dimensional comparison in
\cref{sec:labelfree:three}. Supplementary Section~\SuppPrereg{} lists every criterion of every
run, including the earlier runs of July 2026, with the departures from the plans.

The criteria and their numbers are those of the pre-registration documents. Gates
decide whether a study proceeds to its next stage; kill conditions withdraw a claim when
triggered. In the gate G2 of the three-dimensional comparison, (a) is a plausibility check of
the supervised side, (b) the accuracy of the label-free network and (c) its transfer to the finer mesh. The amendment of the gate, written after the initial
run, assessed (a) from 64 labels only; the naive baselines of (a) are the nearest-neighbour field and the scaled
polynomial of \cref{sec:methods:networks}. In the statements, $\mathcal{L}_D$ and
$\mathcal{L}_K$ are the losses \eqref{eq:ld} and \eqref{eq:lk}.

The noise guard compares a network with a comparator network over three seeds each, through
\begin{equation}
\begin{split}
\mathrm{rel}&=\frac{\mathrm{mean}(\text{network})}{\mathrm{mean}(\text{comparator})}-1,\\
\mathrm{SE}_{\mathrm{rel}}&=\frac{\bigl(\mathrm{sd}(\text{network})^2/3
+\mathrm{sd}(\text{comparator})^2/3\bigr)^{1/2}}{\mathrm{mean}(\text{comparator})},
\end{split}
\label{eq:guard}
\end{equation}
where $\mathrm{mean}$ and $\mathrm{sd}$ are the mean and the sample standard deviation of the metric over a network's seeds. The network is lower beyond the guard when $\mathrm{rel}$ is below minus the guard,
$\max(10\%,\,2\,\mathrm{SE}_{\mathrm{rel}})$, and higher beyond it when $\mathrm{rel}$
exceeds the guard; within it no difference is shown, which is not equivalence. The
three-dimensional energy-norm run applies it to each seed's median over the instances.

\begingroup\footnotesize\setlength{\LTcapwidth}{\linewidth}
\begin{longtable}{@{}>{\raggedright\arraybackslash}p{0.145\linewidth}>{\raggedright\arraybackslash}p{0.405\linewidth}>{\raggedright\arraybackslash}p{0.21\linewidth}>{\raggedright\arraybackslash}p{0.12\linewidth}@{}}
\caption{The pre-registered claims of the comparisons reported in this paper, with the outcomes that the code recorded at the time.}\label{tab:claims}\\
\toprule
Run & Criterion & Measured & Outcome \\
\midrule
\endfirsthead
\caption[]{The pre-registered claims of the comparisons reported in this paper, with the outcomes that the code recorded at the time. (continued)}\\
\toprule
Run & Criterion & Measured & Outcome \\
\midrule
\endhead
\bottomrule
\endfoot
July run (2D), 16 July 2026 & K5: conjugate gradients from the prediction of a label-free network (64 validation instances) save $<$ 20\% of the iterations from zero & $-$0.9\% & triggered \\
\midrule
Three-dimensional comparison, 28 September 2026 & G2 (a), budgets $\geq$ 64 (amendment): displacement-supervised transformer with the energy term at least 3 times better than the zero field in displacement, and better than both naive baselines & 3.9 / 12.0 / 21.3 times; naive beaten & met \\
 & G2 (a), every budget (the registered form, reported for reference) & 2.71 times at 16 labels & not met \\
 & G2 (b): label-free displacement error within +10\% of the displacement-supervised transformer's at 1,024 labels, and relative energy-gap advantage $\geq$ 40\% at every budget & $-$36.3\%; 99.95\% / 99.85\% / 98.86\% / 99.97\% & met \\
 & G2 (c): finer mesh, zero-shot displacement error $\leq$ 1.25 times in-band, and better than both naive baselines & 8.62 times; best naive 0.401 against 0.258 & not met \\
 & Gate G2 = (a) and ((b) or (c)), amended &  & GO \\
 & Gate G2, registered form (reference) &  & NO-GO \\
 & KP4: finer mesh, zero-shot displacement error $>$ 1.5 times in-band, or a naive baseline better & 8.62 times & triggered \\
\midrule
Two-dimensional comparison, 9 October 2026 & H1: label-free transformer lower than the displacement-supervised transformer ($\mathcal{L}_D$) in relative energy gap beyond the noise guard, 1,024 labels & $-$71.3\% (guard 12.4\%) & supported \\
 & H2a: energy-norm-supervised transformer ($\mathcal{L}_K$) lower than the displacement-supervised transformer ($\mathcal{L}_D$) in relative energy gap beyond the guard & $-$78.2\% (guard 16.4\%) & supported \\
 & H2b: energy-norm-supervised transformer lower than the displacement-supervised transformer in von Mises error beyond the guard & $-$54.9\% (guard 12.3\%) & supported \\
 & H3 (a reading, no criterion): energy-norm-supervised transformer against the label-free transformer in relative energy gap & $-$24.2\% (guard 38.1\%) & no difference shown \\
\midrule
Three-dimensional energy-norm run, 25 October 2026 & H2a-3D: energy-norm-supervised transformer ($\mathcal{L}_K$) lower than the displacement-supervised transformer ($\mathcal{L}_D$) in relative energy gap beyond the noise guard, on the seeds' medians & $-$98.2\% (guard 11.7\%) & supported \\
 & H2b-3D: the same in von Mises error & $-$83.6\% (guard 10.0\%) & supported \\
 & H3-3D (a reading without a criterion): energy-norm-supervised transformer against the label-free transformer in relative energy gap, on the seeds' medians & $-$41.0\% (guard 27.0\%) & energy-norm-supervised transformer lower \\
\end{longtable}
\endgroup
\par\medskip

\section*{CRediT authorship contribution statement}

Ruifeng Cao: Conceptualization, Data curation, Formal analysis, Methodology, Project
administration, Software, Validation, Visualization, Writing -- original draft, Writing --
review \& editing. Xidan Song: Data curation, Investigation, Resources, Validation,
Writing -- review \& editing.

\section*{Declaration of competing interest}

The authors declare the following financial interests/personal relationships which may be
considered as potential competing interests: Ruifeng Cao holds the copyright in the code
released with this article and offers licences for commercial use of the code and of the
trained networks, both of which are licensed for non-commercial use only.

\section*{Funding}

This research did not receive any specific grant from funding agencies in the public,
commercial, or not-for-profit sectors.

\section*{Data availability}

{\raggedright\emergencystretch=3em The code, configurations and records listed in
supplementary Section~\SuppRepro{} are available at
\url{https://github.com/RuifengCao/label-free-fe-surrogates}. The version described here is
tagged \texttt{statics-paper-1.0} and archived at Zenodo \cite{Cao2026Zenodo},
\url{https://doi.org/10.5281/zenodo.23815187}. The repository as it stood on 11 October 2026,
when it moved to its present address, is archived with its full history by Software Heritage
as the snapshot
\href{https://archive.softwareheritage.org/swh:1:snp:79f91af6d07e45a45933743bda38253b49c82bb6}{\nolinkurl{swh:1:snp:79f91af6d07e45a45933743bda38253b49c82bb6}}.\par}

The code is licensed under the PolyForm Noncommercial License 1.0.0, which permits its
use, modification and redistribution for any non-commercial purpose, including research
and teaching; licences for commercial use are available from the first author. The records
and the pre-registration documents are licensed under CC BY 4.0.

Of the networks trained in the runs reported here, those whose states the records and the
repository's provenance note name with their SHA-256 hashes are archived at Zenodo
\cite{Cao2026Networks}, \url{https://doi.org/10.5281/zenodo.23815225}, under CC BY-NC 4.0, and
licences for their commercial use are also available from the first author. The others are
not archived and can be retrained from the configurations, seeds and versions of the code
that the repository records.

{\emergencystretch=3em\leavevmode\raggedright The validation sets, training pools and holdout
sets of both corpora and the finer set are archived at Zenodo \cite{Cao2026Data},
\url{https://doi.org/10.5281/zenodo.24091113}, under CC BY 4.0. They hold the meshes,
loads, constraints and reference solutions, exported from the files that the runs read,
whose SHA-256 hashes the archive records. The stiffness matrices are omitted; the archived
code rebuilds them. The other instances of the corpora, and those generated only for
readings taken after the runs, are not archived. The same code regenerates them, though
not necessarily bit for bit, since a mesh can differ slightly between machines and between
versions of the mesh generator.\par}

\section*{Declaration of generative AI and AI-assisted technologies in the manuscript preparation process}

During the preparation of this work, the authors used Claude Fable 5, Claude Fable 5.1 and
Claude Opus 5.5 (Anthropic) in order to draft and revise the text, to write, test and
review the code that runs the experiments and generates the tables, figures and quoted
numbers, to write the TikZ code that draws the graphical abstract, and to check the text
against the records. No image was generated or altered by these tools: the figures, and the
stress fields in the graphical abstract, are drawn from the records by that code, and the
rest of the graphical abstract by the TikZ code. After using these tools, the authors
reviewed and
edited the content as needed and take full responsibility for the content of the published
article.

\clearpage
\includepdf[pages=-]{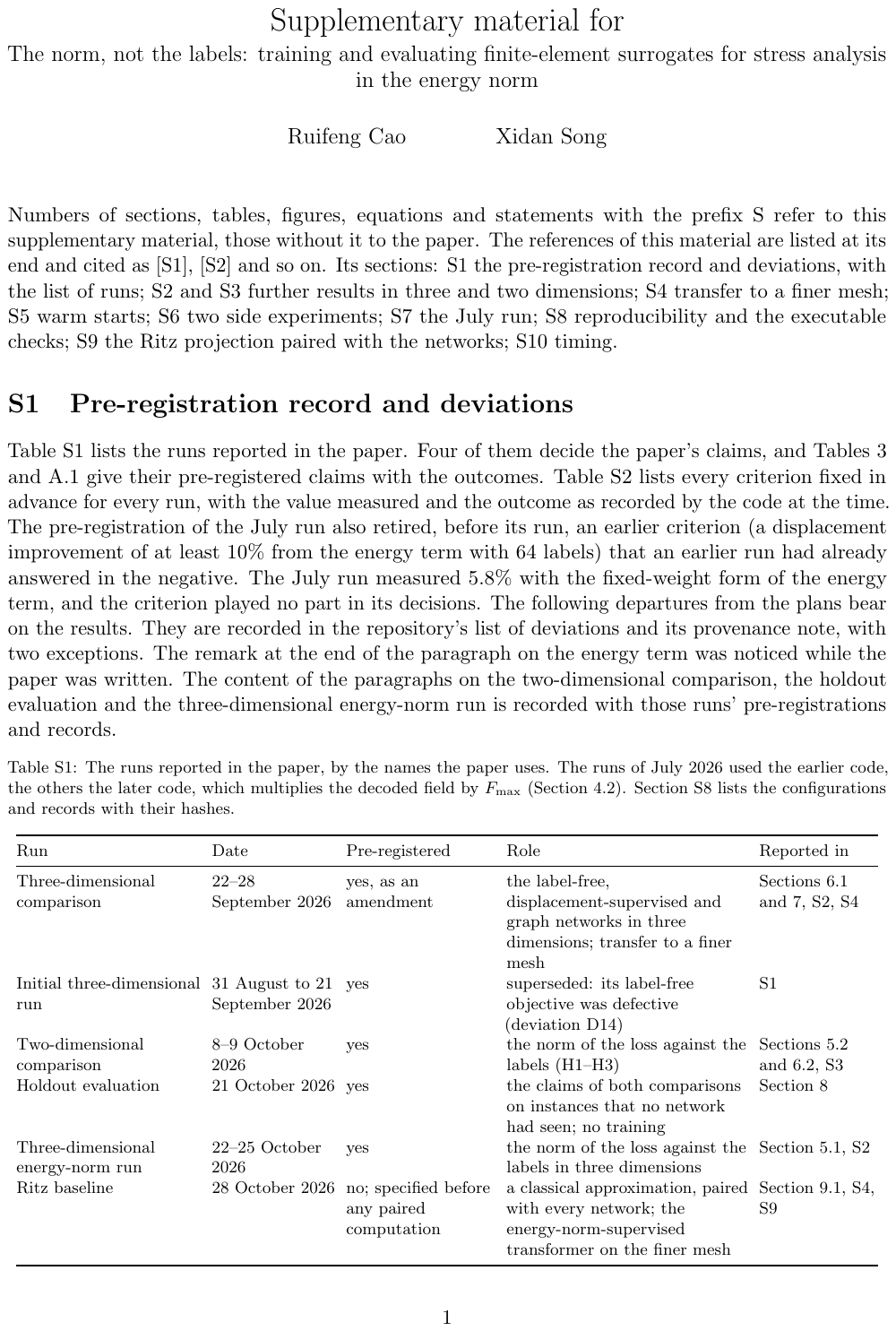}
\end{document}